\documentstyle[preprint,aps,prb,epsf]{revtex}
\begin{document}
\tightenlines
\draft
\title{
Vertex-corrected perturbation theory for the electron-phonon problem 
with non-constant density of states
}
\author{J. K. Freericks }
\address{
Department of Physics, Georgetown University, Washington, DC
20057}
\author{ V. Zlati\'c}
\address{Institute of Physics, 10000 Zagreb, Croatia}
\author{Woonki Chung}
\address{
Department of Physics, Georgetown University, Washington, DC
20057}
\author{ Mark Jarrell}
\address{Department of Physics, University of Cincinnati, Cincinnati, OH 45221}
\date{\today}
\maketitle
\widetext
\begin{abstract}
A series of weak-coupling perturbation theories which include the 
lowest-order vertex corrections are applied to the 
attractive Holstein model in infinite dimensions. 
The approximations are chosen to reproduce the iterated perturbation theory 
in the limit of half-filling and large phonon frequency (where the Holstein 
model maps onto the Hubbard model). Comparison is made with quantum Monte 
Carlo solutions to test the accuracy of different approximation schemes. 
\end{abstract}
\section{Introduction}

The theory of superconductivity, as first developed by Bardeen, Cooper, and 
Schrieffer,\cite{bcs} and then generalized to strong coupling
by Migdal,\cite{migdal} Eliashberg,\cite{eliashberg}, 
and others\cite{parks,reviews}, has proven to be one of the most accurate 
theories of solid-state physics.  Properties of conventional low-$T_c$
materials are typically explained to accuracies of one percent or better.
Newly discovered materials, however, which have moderate $T_c$'s,
do not fit into the parameter regimes studied so successfully
in the 50's and 60's.
These materials, such as the A15 compounds, Ba$_{1-x}$K$_x$BiO$_3$, and
K$_3$C$_{60}$ have large phonon energy scales relative to the inverse electronic
density of states, so that both the effect of the energy dependence of the
bare electronic density of states (i.~e., nonconstant density of states), and 
the effect of vertex corrections may become important in their description.
We examine here a series of different weak-coupling perturbation theories that
include both the
effects of nonconstant density of states and of vertex corrections
to ascertain what methods should be used for real materials calculations
of these higher $T_c$ compounds.  Since the most popular implementation of
Migdal-Eliashberg theory is manifestly particle-hole symmetric (because of
the neglect of the energy dependence of the electronic density of states)
it is not clear what the most accurate approximation scheme is when the
electron filling is doped away from half-filling and the energy dependence of
the electronic density of states becomes important.
Our strategy is to solve a model system
(the Holstein model) in the infinite-dimensional limit via the dynamical
mean field theory.  This allows us to compare numerically
exact quantum Monte Carlo 
solutions (in the thermodynamic limit) with the different perturbative
approximations.

The Holstein model\cite{holstein} consists of conduction electrons  that
interact with local (Einstein) phonons:
\begin{equation}
  H = - \frac{t^*}{2\sqrt{d}} \sum_{\langle j,k\rangle \sigma} ( 
   c_{j\sigma}^{\dag }
  c_{k\sigma} + c_{k\sigma}^{\dag  } c_{j\sigma} ) + \sum_j  (gx_j-\mu)
  (n_{j\uparrow}+n_{j\downarrow}) + \frac{1}{2} M \Omega^2 \sum_j x_{j}^2 +
  {{1}\over{2}} \sum_j \frac{p_{j}^2}{M}.
\label{eq: holham}
\end{equation}
In Eq.~(\ref{eq: holham}),
$c_{j\sigma}^{\dag}$ ($c_{j\sigma}$) creates (destroys) an electron at
site $j$ with spin $\sigma$, $n_{j\sigma}=c_{j\sigma}^{\dag}c_{j\sigma}$ is
the electron number operator, and $x_j$ ($p_j$) is the phonon coordinate
(momentum) at site $j$.  The hopping is chosen to be between the nearest
neighbors $(j$ and $k)$ of a hypercubic lattice in $d$-dimensions and the 
unit of energy is the rescaled matrix element $t^*$ (so that the kinetic
energy remains
finite as $d\rightarrow\infty$).  The phonon has a mass $M$ (chosen to
be $M=1$), a
frequency $\Omega$, and a spring constant $\kappa\equiv M\Omega^2$ associated
with it.  The deformation potential is denoted by $g$ and it governs the
strength of the coupling of electrons to phonons. The
effective electron-electron interaction strength (mediated by the phonons)
is then the bipolaron binding energy
\begin{equation}
  U\equiv - \frac{g^2}{M\Omega^2}=-\frac{g^2}{\kappa} .  
 \label{eq: udef}
\end{equation}
A chemical potential $\mu$ is employed to adjust the total electron filling
with $\mu=U$ corresponding to half-filling in the exact solution.

In the instantaneous limit where $U$ remains finite and $g$ and $\Omega$ are
large compared to the bandwidth $(g,\Omega\rightarrow\infty ,U={\rm
finite})$, the Holstein model maps onto the attractive Hubbard 
model\cite{hubbard}
\begin{equation}
  H = - \frac{t^*}{2\sqrt{d}} \sum_{\langle j,k\rangle \sigma} ( 
   c_{j\sigma}^{\dag }
  c_{k\sigma} + c_{k\sigma}^{\dag  } c_{j\sigma} ) -(\mu-\frac{U}{2})
\sum_j (n_{j\uparrow}+
n_{j\downarrow}) + U \sum_j   n_{j\uparrow}n_{j\downarrow},
\label{eq: hubbham}
\end{equation}
with $U$ defined by Eq.~(\ref{eq: udef}).

Two cases of the Holstein Hamiltonian have well-established perturbative
expansions.  In the limit where the phonon frequency becomes small $\Omega
\rightarrow 0$, but $U$ remains finite, Migdal-Eliashberg 
theory\cite{migdal,eliashberg}
is known to be an accurate approximation.  Migdal-Eliashberg theory is
a self-consistent Hartree-Fock approximation that employs fully dressed
phonon propagators and neglects all vertex
corrections.  Typically, the energy dependence of the electronic
density of states is also neglected, so the theory is evaluated with
a constant density of states. (This latter approximation is always particle-hole
symmetric and maps onto the limit of half-filling when the nonconstant
density of states is used.)  The second limit is the large phonon
frequency limit $\Omega\rightarrow\infty$ with $U$ also remaining finite.
In this case, a truncated perturbative expansion through second 
order\cite{georges_kotliar,freericks_ipt}
(which includes the lowest-order vertex correction) is known to be accurate
at half-filling for a large range of $U$ values, because it properly reproduces
both the weak-coupling and strong-coupling limits of the Hubbard model.
It would be nice to construct an approximation scheme that continuously connects
these two limits as the phonon frequency is varied. However,
no simple approximation
can be found, because the set of diagrams that dress the phonon propagator
in the small-$\Omega$ limit do not correctly dress the phonon propagator in
the large-$\Omega$ limit; in the 
large-frequency limit the interaction is only between up and down spins, but in
the small-frequency limit the interaction is between all spins.  But, if
one is willing to examine perturbative expansions that are truncated
with respect to the fluctuating diagrams, then approximations can be 
constructed that agree with the two known limits through the order
of diagrams included in the expansion.  The description of these different 
methods is both subtle and technical, and will be covered in detail in Section
II.  Here we only want to comment that the previous calculations for the
electron-phonon problem that were called the iterated perturbation 
theory\cite{freericks_ipt} (IPT)
are actually based on a truncated perturbation theory about the Hartree
mean-field theory solution, which does not reproduce the Migdal-Eliashberg
(Hartree-Fock)
limit properly.  A more promising approach for all $\Omega$
is to construct a truncated
perturbation theory about the Hartree-Fock mean-field theory solution, which
is done here. In addition, we also examine some simple methods that can be
used to repair inconsistencies that develop in the IPT as the system is 
doped off of half-filling.  This method entails self-consistently renormalizing
the higher-order fluctuations of the Hartree or Fock 
diagrams to all orders, which allows for the electron filling
on the lattice to be different from the electron filling of the mean-field
theory solution when the higher-order fluctuations are included. Unfortunately, 
we find that all attempts to produce an accurate perturbation theory that 
reproduces the IPT in the instantaneous limit are not very accurate for
moderate phonon frequencies.

These perturbative approaches, and the many-body problem in general,
simplify when the infinite-dimensional limit\cite{metzner_vollhardt}
is taken $(d\rightarrow\infty)$. Then the many-body problem
becomes a local problem that retains its complicated dynamics in
time.  The hopping integral is scaled to zero in such a fashion that the
free-electron kinetic energy remains finite while the self energy
for the single-particle Green's function and the irreducible vertex functions
have no momentum dependence and are functionals of the
local Green's function\cite{metzner_vollhardt,schweitzer_czycholl,metzner}.
This limit retains the strong-correlation effects that arise from
trying to simultaneously minimize both the kinetic energy 
and the potential energy, and hence has relevance for three-dimensional
materials.

Of course, we can also solve the infinite-dimensional
Holstein model using a quantum Monte Carlo
method, which contains all effects due to phonon renormalizations,
vertex corrections, and nonconstant electronic density of states.
We use these solutions as a benchmark to test the accuracy of the
different approximation methods and to determine what is the most fruitful
approximation for the electron-phonon problem.

We employ a Green's function formalism to solve the many-body problem.
The local Green's function is defined to be
\begin{equation}
G_{loc}(i\omega_n)\equiv -\int_0^{\beta} d\tau e^{i\omega_n\tau}
{{\rm Tr} \langle e^{-\beta H}T_{\tau} c(\tau)
c^{\dag}(0)\rangle\over {\rm Tr} \langle e^{-\beta H}\rangle } ,
\label{eq: greendef}
\end{equation}
and is calculated directly from a self-consistent quantum Monte Carlo
procedure described elsewhere.\cite{qmc_holstein}
Static two-particle properties can also be determined since the 
irreducible vertex function is local\cite{zlatic_horvatic_infd}.  The 
static susceptibility for charge-density-wave (CDW) order is given by
\begin{eqnarray}
\chi^{CDW}({\bf q})&=& {1\over2N}\sum_{{\bf R}_j-{\bf R}_k\sigma\sigma'}
e^{i{\bf q}\cdot
({\bf R}_j-{\bf R}_k)} T\int_{0}^{\beta} d\tau \int_{0}^{\beta} d\tau '
[\langle n_{j\sigma}(\tau) n_{k\sigma'}(\tau ')\rangle - 
\langle n_{j\sigma}(\tau)\rangle\langle n_{k\sigma'}
(\tau ')\rangle ],\cr
&=& T\sum_{mn} \tilde\chi_{mn}^{CDW} ({\bf q}),
\label{eq: chicdw}
\end{eqnarray}
at each ordering wavevector ${\bf q}$ [the indices $m$ and $n$ denote
Matsubara frequencies $i\omega_n=i\pi T(2n+1)$].  Dyson's equation for the 
two-particle Green's function becomes\cite{qmc_holstein,zlatic_horvatic_infd}
\begin{equation}
\tilde\chi_{mn}^{CDW}({\bf q})=\tilde\chi_{m}^0({\bf q})\delta_{mn}
-T\sum_p \tilde\chi_m^0({\bf q})\Gamma_{mp}^{CDW}\tilde
\chi_{pn}^{CDW}({\bf q}) ,
\label{eq: cdwdys}
\end{equation}
with $\Gamma_{mn}^{CDW}$ the irreducible vertex function in the CDW
channel.  

The
bare CDW susceptibility $\tilde\chi_n^0({\bf q})$ in 
Eq.~(\ref{eq: cdwdys}) is defined in terms of the {\it dressed}
single-particle Green's function
\begin{eqnarray}
\tilde\chi_n^0({\bf q})&\equiv&-{1\over N} \sum_{\bf k} G_n({\bf
k})G_n({\bf k+q})\cr
&=&-\frac{1}{\pi}\int_{-\infty}^{\infty}
dy {{e^{-y^2}}\over{i\omega_n+\mu-\Sigma_n-y}}\int_{-\infty}^{\infty}dz
\frac{e^{-z^2}}{i\omega_n+
\mu-\Sigma_n-X({\bf q})y-z\sqrt{1-X^2({\bf q})}}
\label{eq: chi0cdw}
\end{eqnarray}
with all of the wavevector dependence described by the
scalar\cite{brandt_mielsch,muellerhartmann} $X({\bf q})
\equiv \sum\nolimits_{j=1}^d \cos {\bf q}_j/d$.
The integral for $\tilde\chi_m^0(X)$
in Eq.~(\ref{eq: chi0cdw}) can then be
performed analytically\cite{brandt_mielsch} for the ``checkerboard'' CDW phase
$\tilde\chi_n^0(X=-1)=-{{G_n}/({i\omega_n+\mu-\Sigma_n}})$.
The irreducible vertex function $\Gamma_{mn}^{CDW}$ is either directly
calculated in a perturbative expansion (described below) or
is determined by inverting the Dyson equation in Eq.~(\ref{eq: cdwdys}) for the
local charge susceptibility (QMC).

A similar procedure is used for the singlet $s$-wave superconducting
(SC) channel.  The
corresponding definitions are as follows:  The static susceptibility
in the superconducting channel is defined to be
\begin{equation}
\chi^{SC}({\bf q})\equiv {1\over N}\sum_{{\bf R}_j-{\bf R}_k}
e^{i{\bf q}\cdot
({\bf R}_j-{\bf R}_k)} T\int_{0}^{\beta} d\tau \int_{0}^{\beta} d\tau '
\langle c_{j\uparrow}(\tau)
c_{j\downarrow}(\tau)c_{k\downarrow}^{\dag}(\tau ')c_{k\uparrow}^{\dag}(\tau ')
\rangle 
= T\sum_{mn} \tilde\chi_{mn}^{SC} ({\bf q}),
\label{eq: chisc}
\end{equation}
for superconducting pairs that carry momentum ${\bf q}$;  Dyson's equation
becomes
\begin{equation}
\tilde\chi_{mn}^{SC}({\bf q})=\tilde\chi_{m}^0{'}({\bf q})\delta_{mn}
-T\sum_p \tilde\chi_m^0{'}({\bf q})\Gamma_{mp}^{SC}\tilde
\chi_{pn}^{SC}({\bf q}) ,
\label{eq: scdys}
\end{equation}
with $\Gamma_{mn}^{SC}$ the corresponding irreducible vertex function to 
describe the SC channel; the bare pair-field susceptibility becomes
\begin{eqnarray}
\tilde\chi_n^0{'}({\bf q})&\equiv&{1\over N} \sum_{\bf k} G_n({\bf
k})G_{-n-1}({\bf -k+q})\cr
&=&\frac{1}{\pi}\int_{-\infty}^{\infty}
dy {{e^{-y^2}}\over{i\omega_n+\mu-\Sigma_n-y}}\int_{-\infty}^{\infty}dz
\frac{e^{-z^2}}{i\omega_{-n-1}+
\mu-\Sigma_{-n-1}-X({\bf q})y-z\sqrt{1-X^2({\bf q})}}
\label{eq: chi0sc}
\end{eqnarray}
with the special value 
$\tilde\chi_n^0{'}(X=1)=-{\rm Im}G_n/{\rm Im}(i\omega_{n}-\Sigma_{n})$ for the
SC pair that carries no net momentum; and finally the irreducible vertex 
function is also  either directly
calculated in a perturbative expansion (described below) or
is determined by inverting the Dyson equation in Eq.~(\ref{eq: scdys}) 
for the local pair-field susceptibility (QMC).

The transition temperature of the infinite-dimensional Holstein 
model is then found by calculating the temperature at which the relevant 
susceptibility diverges (CDW or SC).
This transition temperature is found by locating
the temperature where the scattering matrix (in the relevant
channel)
\begin{equation}
T_{mn}=-T\Gamma_{mn}\tilde\chi_n^0,
\label{eq: Tmat}
\end{equation}
has unit eigenvalue\cite{owen_scalapino} (note that the local Green's 
functions are always used in the evaluation of the bare susceptibility
$\tilde\chi_0$).  

The remainder of the paper is arranged as follows:
In Section II we describe the formalism employed to derive the different 
truncated perturbative expansions for both the electronic self energy
and for the irreducible vertex functions.  Particular attention is paid
to the mapping from the lattice problem onto an effective impurity problem,
and how to extract a perturbative expansion that includes the lowest-order
fluctuations beyond a mean-field theory solution.  In Section III we describe
the details of the calculational procedure and present our results in Section 
IV.  Section V contains a summary and our conclusions.

\section{Lattice-impurity mapping}
The properties of the Holstein model are calculated by 
mapping the lattice problem onto the single-impurity 
Wolff-Holstein model\cite{wolff,holstein}, and solving the impurity self 
energy by a truncated perturbation expansion.
To define the mapping, we begin with the self-energy of the Holstein model, 
which is momentum independent in infinite dimensions, and write the local 
Green's function for the lattice as, 
\begin{equation}
                                  \label{G_loc}
G_{loc}(\omega) 
= 
\sum_{\bf p}
\frac 1 
{\omega - (\epsilon_{\bf p}-\mu) 
- 
\Sigma(\omega)}, 
\end{equation}
where $\epsilon_{\bf p}$ is the noninteracting electronic bandstructure
for the lattice (assuming spin degeneracy $n_{\uparrow} = n_{\downarrow}=n/2$).
The local self-energy $\Sigma(\omega)$ is defined by 
the functional, $\Sigma(\omega) = \Sigma\left[G_{loc}\right]$, 
where $\Sigma\left[G_{loc}\right]$ represents the sum of all the 
skeleton diagrams generated by the perturbation theory, with U as the 
expansion parameter (in a Baym-Kadanoff expansion\cite{baym_kadanoff}). 
On the other hand, the same set of skeleton diagrams appears in 
the self-energy of an impurity problem described by the Wolff-Holstein 
Hamiltonian, 
\begin{equation}
                                \label{H_wolff}
H_{imp}= H_0 + gx_0 (n_{0\uparrow} + n_{0\downarrow}) + \frac{p_0^2}{2M}+
\frac{1}{2}M\Omega^2x_0^2,
\end{equation}
where $H_0$ describes a band of non-interacting electrons,
$n_{0\sigma}$ is the number operator 
for conduction electrons of spin $\sigma$ at the impurity site and $x_0$ ($p_0$)
is the ``impurity'' phonon coordinate (momentum).  
The renormalized electron propagator at the impurity site can be written as, 
\begin{equation}
                                \label{G_imp}
G_{imp}(\omega) 
= 
\frac 1 
{G_{0}^{-1}(\omega)
- 
\Sigma(\omega) }.  
\end{equation}
Here  $G_0$ is the free-electron propagator at the impurity site
(which is often called the effective-medium propagator), 
and $\Sigma(\omega)$ 
describes the effects due to the coupling to the local phonon. Note that the
effective-medium $G_0$ is not equal to the noninteracting 
lattice propagator except when $U=0$.

Since the impurity and the lattice self-energy functionals, 
$\Sigma\left[G_{imp}\right]$ and $\Sigma\left[G_{loc}\right]$ are the same,  
the self energies will also be the same if
$G_{imp}(\omega)=G_{loc}(\omega)$.
Thus, the lattice problem maps onto the impurity problem, provided 
the effective medium propagator $G_0$
is adjusted so that the right hand sides of
Eqs.~(\ref{G_loc}) and (\ref{G_imp}) are equal.  
In general, the self-energy functional $\Sigma\left[G_{loc}\right]$ 
is not known. But equivalent
expansions can also be made for the impurity self-energy by rearranging the
skeleton-diagram expansion. 
For example, a functional defined on the effective-medium propagator 
$G_0(\omega)$ can be employed, such that 
$\Sigma(\omega)=\Sigma_0\left[G_0\right]$, 
where $\Sigma_0\left[G_0\right]$ represents the sum of all connected 
graphs generated by Wick's theorem (without any resummations). 
If the graphs for $\Sigma\left[G_{loc}\right]$ and $\Sigma_0\left[G_0\right]$
are summed to all orders, then the self energies must agree. Furthermore,  
the effective medium $G_0(\omega)$ can be uniquely determined by either 
$G_{0}^{-1}(\omega)=G_{loc}^{-1}(\omega)+\Sigma\left[G\right]$
or 
$G_{0}^{-1}(\omega)=G_{loc}^{-1}(\omega)+ \Sigma_0\left[G_0\right]$. 

Since the exact self-energy for the Wolff-Holstein model  
with an arbitrary density of states is not known, 
our strategy is to make a truncated expansion for the 
impurity self-energy functional 
$\Sigma_{app}\left[G_0\right]$, and 
define the self-energy for the lattice problem via  
$\Sigma(\omega)=\Sigma_{app}\left[G_0\right]$, 
The self consistency is then imposed on $G_0(\omega)$,   
so that $G_{loc}(\omega)=G_{imp}(\omega)$. 
The hope is, that a controlled expansion of the impurity self-energy   
will lead to an accurate solution of the impurity problem, and of the resultant 
lattice problem.  
Here, we use the lowest-order Yosida-Yamada
expansion, which is known to provide reliable answers for the Wolff
model describing magnetic impurities\cite{yosida_yamada,zlatic.89}.

The Yosida-Yamada expansion for the effective impurity problem
(with an unknown density of states) is obtained by 
a partial resummation of diagrams that allows us to construct the  
self-energy functional in terms of diagrams that involve either  
Hartree or Hartree-Fock Green's functions. 
That is, one considers the self-energy corrections with respect 
to the mean-field solution, and rewrites the Dyson equation as, 
\begin{equation}
                                \label{G_YY}
G_{imp}(\omega) 
= 
\frac{1}{[G_{MF}(\omega)]^{-1} - \Sigma_{YY}^{MF}[G_{MF}] }. 
\end{equation}
where $\Sigma_{YY}^{MF}$
is the sum of all self-energy diagrams defined in terms of 
the mean-field propagator for the impurity problem, $G_{MF}$, 
and 
\begin{equation}
                                   \label{G_MF}
G_{MF}^{-1}(\omega)
= G_0^{-1}(\omega)-\Sigma_{MF}. 
\end{equation}
It is clear that functionals $\Sigma_{YY}^{MF}[G_{MF}]$
defined with different mean-field propagators are not the same, and 
it is not known {\it a priori} which truncated
approximation for the impurity self energy 
leads to the most accurate result for the lattice problem.
The mean-field solution has an average electron filling $n_{MF}$ defined
by
\begin{equation}
                                    \label{n_MF} 
n_{MF} 
=
-\frac 2{\pi}\, Im\;
\int_{}^{}  d\omega \; G_{MF}(\omega) f(\omega),   
\end{equation}
with $f(w)=1/[1+\exp(\beta\omega)]$ the Fermi factor (the factor of
2 is from the spin summation).
This filling is usually different from the electron filling on the lattice which
is determined by the same equation, but with $G_{MF}$ replaced by $G_{loc}$.
In this contribution, we consider four different expansions to the
self energy
and compare the corresponding results to the quantum Monte Carlo solution.

Our first approximation is called the Hartree expansion (H) and it includes
a truncated perturbation-theory expansion about the Hartree mean-field solution,
including all diagrams through second order.  This expansion should
not be confused with the Hartree approximation, which does not include any
of the second-order vertex corrections.
These self-energy corrections with respect to the Hartree mean-field solution 
are constructed by using the Hartree self-energy diagram shown in Figure 1, 
to construct the mean-field propagator of Eq.~(\ref{G_YY}). That is, we take 
\begin{equation}
                                   \label{G_hartree}
\Sigma_{MF}=
\Sigma_{H}
= Un_H, 
\end{equation}
which implies that the Hartree self-energy insertions are included to all
orders in the perturbative expansion.
Here $n_H$ is the (Hartree) mean-field particle number defined by 
Eq.~(\ref{n_MF}) with $G_{MF}=G_H$. 
Next, 
the Hartree mean-field Green's function is used in 
evaluating the truncated Yosida-Yamada self-energy functional to determine the 
approximate self-energy.   
This functional is given in Figure 1 through second-order
in $U$ where the electronic propagators (solid lines) are the Hartree mean-field
propagators from Eq.~(\ref{G_MF}), the phonon propagators (wiggly lines) 
are bare $D(\omega)=-1/[M(\Omega^2-\omega^2)]$,
and the impurity self-energy which renormalizes the effective field
$G_0$ in Eqn.(\ref{G_imp})  becomes
\begin{equation}
                                          \label{sigma_hartree_yamada}
\Sigma(\omega)
=
\Sigma_{H}
+
\Sigma_{YY}^H[G_H]. 
\end{equation}
Note that this procedure is summing an infinite class of diagrams (the Hartree
self-energy insertions), but is truncating the perturbation theory
{\it with respect to the Hartree mean-field-theory solution} to include
both the Fock and
the second-order fluctuating terms.  It turns out that this procedure
is identical in form to what was previously called the IPT 
approximation\cite{freericks_ipt}, since
the renormalization of the effective medium by the Hartree self energy (to
construct the Hartree mean-field Green's function) can be absorbed by
a redefinition of the chemical potential (since the Hartree self energy
just provides a frequency-independent shift), with the exception being 
the inclusion of the fifth diagram, which renormalizes the Hartree diagram
by the Fock self-energy insertion.  The omission of the fifth diagram
in the older IPT paper\cite{freericks_ipt} may be thought to be innocuous,
because that diagram vanishes at half-filling and should not affect the
results much off of half filling. However, when the coupling strength becomes
large, it's effects do become strong, as shown in the next
section. To reiterate, the difference between the present expansion and the
older IPT work is that the fluctuating diagrams in $\Sigma_{YY}^H[G_H]$ are
evaluated with $G_H$ which does not include the frequency-independent
shift from the fifth diagram of Figure 1.  The IPT calculation included the
frequency-independent shifts to all orders, and hence evaluated the first
four diagrams of Figure 1 using a different mean-field propagator than $G_H$.
The inclusion of the Fock self-energy
insertion into the Hartree diagram is just one of the subtle, and often
neglected diagrams that needs to be included in a truncated approximation
that includes all diagrams up to a given order.

Our second approximation is called the Hartree-Fock expansion (HF) 
and is obtained by using the Hartree-Fock self-energy 
\begin{equation}
                                   \label{sigma_hartree_fock}
\Sigma_{MF}=
\Sigma_{HF}
=Un_{HF}+\Sigma_{F} 
\end{equation}
to define the mean-field propagator in Eqn.(\ref{G_MF}), 
and including all second-order diagrams with respect to the Hartree-Fock
mean-field solution (once again, this should not be confused with the
Hartree-Fock approximation which does not include any of the second-order
vertex corrections).
The (Hartree-Fock) mean-field particle number $n_{HF}$
is calculated by using $G_{HF}$ 
in Eqn.(\ref{n_MF}), and the Fock self energy $\Sigma_F$
satisfies
\begin{equation}
                                    \label{sigma_F}
\Sigma_{F}(\omega) = g^2 \int_{}^{}  d\epsilon 
G_{HF}(\omega-\epsilon) D(\epsilon)  f(\epsilon),    
\end{equation}
with $D(\epsilon)$ the bare phonon propagator again.
This mean-field solution sums both the Hartree and Fock self-energy insertions
to all orders, and would be identical to Migdal-Eliashberg theory if the 
dressed phonon propagator was employed in Eq.~(\ref{sigma_F}) rather than
the bare propagator.  As described in the introduction, we are forced to
use the bare propagator if we want to reproduce the IPT limit in the
large-phonon-frequency limit.
The Yosida-Yamada functional defined on $G_{HF}$ is given by the 
diagrams in Figure~2 where the solid lines are now Hartree-Fock propagators.
The self-energy correction to $G_0$ for the HF expansion can then
be written as 
\begin{equation}
                                          \label{sigma_HF_yamada}
\Sigma(\omega)
=\Sigma_{HF}+\Sigma_{YY}^{HF}[G_{HF}].
\end{equation}

The self-energy diagrams given in Figs. 1 and 2 are evaluated on the
imaginary axis by using 
the standard rules \{explicit expressions are given in Eq.~(16) of 
Ref.~\onlinecite{freericks_ipt} [and the suitable modifications for the
Hartree-Fock case]\}, 
and the Matsubara summations occurring in 
$\Sigma_{YY}^H[G_{H}]$ and $\Sigma_{YY}^{HF}[G_{HF}]$ 
are performed numerically for both the Hartree and the Hartree-Fock
expansions. 

If the perturbative expansion for the
self-energy is extended to higher-order in U 
we find additional frequency-independent diagrams [which sum up to
$U(n-n_{MF})$] in addition to the frequency-dependent diagrams
(similar to the fifth diagram in Figure~1).   
Here $n$ is the fully renormalized particle number of the lattice which is
calculated by Eq.~(\ref{n_MF}) with $G_{loc}$ replacing $G_{MF}$.  These 
diagrams arise simply
from the fact that in the exact skeleton expansion
the Hartree diagram is evaluated with the fully 
dressed Green's function (yielding $Un$) rather than with the mean-field
Green's function (which yields $Un_{MF})$.  At half-filling, the two fillings 
$n$ and $n_{MF}$ are usually
equal, but they need not be the same (and never are) away from
half filling.
We introduce a new self-energy functional ${\tilde\Sigma}^H_{YY}[G_{H}]$
(that is based on the Yosida-Yamada functionals defined above) that has all
of the higher-order renormalizations of the Hartree diagram removed
from it [as shown, through third-order
in Figure~3~(a)]. Then we have an exact relation 
$\Sigma(\omega) = Un + {\tilde\Sigma}^H_{YY}[G_{H}] $ for the self energy 
$\Sigma(\omega)$ defined in Eq.~(\ref{G_imp}). 
In this contribution,
the two self-energy functionals $\Sigma^H_{YY}$ and $\tilde\Sigma^H_{YY}$ 
differ by one diagram (the fifth diagram in Figure~1).
We can employ this exact relation to formulate the  renormalized-Hartree 
expansions\cite{horvatic.85} (also called the
n-consistent approximation), which enforces this self-consistency 
condition on the renormalized particle number. 
That is, for a given $\mu$, g, $n_{H}$ and ${\tilde\Sigma}^H_{YY}(\omega)$, 
we solve the transcendental equation, 
\begin{equation}
                              \label{n_h}
n
= -\frac 2{\pi}\, Im\;
\sum_{\bf p}\int d\omega 
\frac{f(\omega)} 
{ \omega  -(\epsilon_{\bf p \sigma} - \mu) 
- 
\,Un - {\tilde\Sigma}_{YY}^H(\omega)},
\end{equation}
to determine the impurity electron filling $n$,
and to obtain the correction to the real part of the total self energy. 
For the Anderson and Wolff impurity models, and for the
two-dimensional Hubbard model,
this renormalized-Hartree approach (RH)
expands the region of the validity of the 
perturbation theory, and allows for the description of stronger 
correlations\cite{horvatic.85,n_consistent}. 
The hope is that this renormalized-Hartree scheme will repair some of the 
inaccuracies of the truncated perturbation theories away from half filling.
It is a much simpler approximation to study than a scheme that tries to
interpolate between the weak-coupling and atomic limits\cite{kajuter_kotliar}
(which is significantly
more challenging to formulate for the electron-phonon problem).

A similar approach can be made for summing higher-order diagrams in the
Hartree-Fock expansion.  In this case, however, an n-consistent
approximation, that sums only the frequency-independent diagrams to all orders,
is likely to be less accurate than one that sums the renormalizations
based on both the Hartree and Fock diagrams.  Hence, we form the exact
relation for the self energy of Eq.~(\ref{G_imp})
\begin{equation}
\Sigma(\omega)=Un+g^2\int d\epsilon G_{loc}(\omega-\epsilon)D(\epsilon)
f(\epsilon)+{\tilde\Sigma}^{HF}_{YY}[G_{HF}],
\end{equation}
and employ it to evaluate the renormalized-Hartree-Fock expansion (RHF)
where the functional ${\tilde\Sigma}^{HF}_{YY}(G_{HF})$ is truncated at
second order [see Figure~3(b)].  This approximation is formed by evaluating
both the Hartree and Fock diagrams with the local Green's function (instead
of $G_{HF}$) but the second-order diagrams continue to be evaluated
with $G_{HF}$.  In this sense it is ``halfway'' between a conserving
approximation (which evaluates all diagrams with $G_{loc}$) and the truncated
expansions described above.

The only remaining objects that need to be determined are the irreducible
vertices for the CDW and SC instabilities.  These vertices are calculated
with the mean-field Green's function (either Hartree or Hartree-Fock)
and have an identical functional
form for all different approximation schemes.  The diagrams
included are presented in Figure~4(a) for the CDW vertex and Figure~4(b)
for the SC vertex.  The solid lines are $G_{MF}$ ($G_H$ or $G_{HF}$)
and the wiggly lines are
the bare phonon propagator.  Explicit expressions for these diagrams
on the imaginary axis are given in Eqs.~(17) and (18) of
Ref.~\onlinecite{freericks_ipt} with $G_0$ replaced by $G_{MF}$.
\section{Calculational Methods}
The calculational methods used are straightforward.  The perturbation expansion
is carried out on the imaginary axis.  We employ an iterational
algorithm to solve the self-consistent equations, which is summarized in 
Figure~5:
\begin{enumerate}
\item
     Start with an initial self-energy $\Sigma(\omega)$;
\item
     Use Eqn.(\ref{G_loc}) to calculate $G_{loc}$;
\item
     Calculate the effective-medium propagator $G_0$ using Eqn.(\ref{G_imp})
     with $G_{imp}=G_{loc}$;
\item
     Calculate the mean-field propagator using Eqn.(\ref{G_MF});
\item 
     Calculate the Yosida-Yamada
     self-energy diagrams as defined by Fig.~1 or Fig.~2, 
     with and without the renormalization of the first-order
     diagrams (as shown in Fig.~3), to determine the total self-energy;
\item
     If the calculation 
for the self-energy 
 has not converged, then adjust the chemical 
     potential to produce the target electron filling; 
\item
     Repeat steps 2-6 until the calculation has converged;
\item
     Once converged, calculate the irreducible vertex functions  from Fig.~4
     and determine 
     the maximal eigenvalue of the scattering matrix in both the CDW and SC 
     channels from Eq.~(\ref{eq: Tmat});
\item
     Repeat 1-8 for another temperature $T$ until the transition temperature
     $T_c$ is found. 
\end{enumerate}
We use an energy cutoff of 256 Matsubara frequencies, which provides accurate
results for temperatures $T$ larger than $0.01t^*$, and our convergence 
criterion is that the Green's functions do not change by more than one part
in $10^9$ from one iteration to the next.  We find that the perturbation
theory typically converges in approximately 100 iterations of the 
main ring in Figure 5.  Sometimes the iterated equations develop
limit cycles, whose oscillations can be suppressed by employing standard
damping methods that average the $i-1$st and $i$th iterations to produce the
starting point for the next iteration.  We also found that if the coupling
strength is large enough, then some of the approximate theories
will have multiple
solutions for $n(\mu)$ near half filling.  In this case, the symmetry point,
with $\mu=U$ becomes unstable (i.~e., the compressibility is negative), and
the Green's functions are no longer purely 
imaginary when the electron filling is
half-filled.  This latter result is an indication of the breakdown of
the approximation method at such a large coupling strength.

There also are some technical details that need to be discussed about the 
quantum Monte Carlo simulations\cite{qmc_holstein}.  The results for the 
transition temperatures
were calculated with a $\Delta \tau=0.4$, and sometimes also with a
$\Delta\tau=0.2$ and then extrapolated to $\Delta\tau=0$, when the
correlations were large enough that the Trotter error was noticeable.
The self-energies and irreducible vertex functions were calculated
with a fixed $\Delta\tau=0.1$ to ensure a high accuracy.  In general, there
are three sources of error to the quantum Monte Carlo simulation:
(i) statistical error; (ii) iteration error; and (iii) systematic error.
The statistical error is the easiest error to control, and is the smallest
of the three sources of error.  The iteration error arises from performing
calculations with the wrong dynamical mean-field $G_0$, because the
calculation has not yet converged fully.  We typically iterate 8 or 9 times,
which provides good convergence for the iterations.  The systematic
error is more difficult to control.  It arises from the Trotter error, and
from other (potentially unknown) sources.  One surprising source of error
arose from the choice of global updating moves for the phonon coordinate.
Global moves that shift the phonon coordinate at every time slice by
an amount randomly chosen between $-dx/2$ and $dx/2$ were supplemented
by global shifts by an amount ranging between $\pm 2g/(m\Omega^2)-dx/2$ and
and $\pm 2g/(m\Omega^2)+dx/2$.  These latter moves were chosen to allow
the phonon coordinate to shift between the two minima of the effective 
phonon potential (separated by $2g/[m\Omega^2]$), which form when the system 
enters the strong-coupling
region, were preformed pairs exist above $T_c$.  When one is in the
weak-coupling regime, where the effective phonon potential possesses a
single-well structure, we find that the self-energy can differ by a few
percent depending on whether the global moves are all chosen randomly, or
if the supplemental global shifts are chosen to move the phonon coordinate
to the ``other'' well (i.e. if the phonon coordinates lie in the right well
then the global shift is chosen to move the system to the left, and vice
versa).  The accuracy was best when the global moves were chosen completely
randomly, which is the method used here.  These results differ by a few percent 
from those shown previously\cite{freericks_ipt}, where the global moves were 
coupled
to the values of the phonon coordinates, and the Trotter time slice was larger.

\section{Results and discussion}
The parameters of the Hamiltonian in Eq.~(\ref{eq: holham}) are chosen to agree
with previous theoretical work, and to represent a parameter range where 
the vertex corrections can cause large effects.  As such, we choose the
phonon frequency to be on the order of
one tenth of the bandwidth $\Omega=0.5t^*$ (with the effective bandwidth
of the Gaussian being about two standard deviations above and below the
center, or approximately $4t^*$).  Most of our calculations 
are performed for three different values of the interaction
strength (i) $g=0.4$ ($U=-0.64$, one sixth of the bandwidth) which corresponds 
to a fairly weak interaction where
there are no preformed pairs, but the transition temperatures are large
enough that the phase diagram can be determined reliably; (ii) $g=0.5$ 
($U=-1.0$, one fourth of the bandwidth) which is in the moderate interaction
regime; and (iii) $g=0.625$ ($U=-1.5625$, two-fifths of the bandwidth)
which is where the system enters the strongly coupled regime,
where numerous preformed pairs are present above $T_c$.

The quasiparticle renormalization factor, the irreducible CDW vertex function,
and the irreducible SC vertex function are plotted in Figs.~6(a-c). 
The vertex functions are averaged to show just the symmetric frequency
component, since that is all that contributes to the eigenvalue of the
scattering matrix for the CDW at half filling, or for the SC at any filling
(this is so is because the eigenvector with maximal eigenvalue
is symmetric with respect to a change in sign of the Matsubara frequency).
The coupling strength is $g=0.4$ and the filling is $n=1.0$,
which yields a bare electron-phonon
coupling strength of $|U|=0.64$, or $\lambda=\rho(0)|U|=0.36$, which would
naively be viewed as quite weak coupling.  However, this is not the case,
as one can estimate the renormalized value of $\lambda$ from $Z(0)-1$ which
lies at about 0.7.  These calculations indicate that higher-order diagrams,
such as those that fully dress the phonon propagator, are important even
when the bare coupling has $\lambda=0.36$.  What is interesting, is that
the perturbative results seem to have the right shape, but need to have the
frequency axis moved to the right, to line up with the Monte Carlo data.
Furthermore, since the perturbation theory underestimates both the
renormalization factor (which will enhance $T_c$) and the vertex function
(which will reduce $T_c$) one might expect these results to tend to cancel
each other out when calculating a phase diagram, which is indeed what
we find below.  Figs.~7(a-c) show the same results for the stronger coupling 
case $g=0.5$ and $n=1.0$ ($|U|=1.0$ and $\lambda=0.56$).  We can see that the
error in the renormalization factor has grown, even though all four 
perturbative methods yield similar values.  The CDW vertex becomes more 
attractive at moderate coupling, which is missed by the perturbation
theory.  The SC vertex has the right qualitative shape, but is underestimated
at low frequencies.  When the coupling strength is increased to $g=0.625$
(not shown) the situation becomes even worse, with the same qualitative
features seen, a growing difference in the renormalization factors,
and an underestimation of the size of the vertices.

We also include, in Fig.~8, a result at quarter filling $n=0.5$ and $g=0.4$ at 
the same temperature $T=1/16$ as in Fig.~6.  One might have thought that the
quarter-filled case would be approximated much better by the perturbation
theory because the effective coupling strength is lower due to the 
reduction of the electronic density of states at the Fermi level, but
we see that the shape of the curves is both qualitatively and quantitatively 
similar for Figs.~6 and 8.  The quantitative agreement has improved, but the 
improvement is not dramatic.  What is surprising is that the calculated
transition temperature agrees much better with the QMC here than at
half filling, because of the cancellation of the errors to both the
self energy and the vertex function.  Note that we only plot the SC vertex
here, because an asymmetric average is needed for the physically relevant
piece of the CDW vertex.

In addition to calculating the single-particle and two-particle self energies,
we can also investigate the phase diagrams of the Holstein model.  Based
on our results for the self energy and the vertex functions, it does not appear
likely that any approximation will be too accurate for the phase diagrams,
but as theorized above,
it is possible that the inaccuracies will cancel out to produce more accurate
results for the transition temperatures.  Such behavior should be viewed
as ``accidental'' agreement with the exact result.

We begin by calculating the transition temperature to the commensurate
(checkerboard) CDW insulator at 
half-filling as a function of the interaction strength.  The QMC 
solution\cite{qmc_holstein}
showed that the maximal $T_c$ was on the order of 1/25th of the bandwidth,
occurring when $U$ was about two-fifths of the bandwidth.  The results
for each of the four expansions shown here is presented in Figure~9.
The horizontal axis is the interaction strength and the vertical axis is the 
transition temperature.  One can see that 
the peak in $T_c$ versus $U$ is not produced by the perturbation
theory, rather the transition temperature continues to increase.
This differs from what happened in the so-called IPT approximation
where the approximate curve for $T_c$ did show a peak.\cite{freericks_ipt}
The Hartree and renormalized-Hartree expansions agree with each other
and the IPT approximation
until $U$ is approximately $-1$, where they differ from each other because
the thermodynamically stable H and RH expansions are no longer at the
particle-hole symmetric point with $\mu=U$, but rather have different values
of $\mu$ (and no longer have purely imaginary Green's functions when evaluated
along the
imaginary axis).  This is the point where the kink appears in the $T_c(g)$
curve.  The remarkable agreement of the IPT approximation in
predicting both the peak position and the peak height accurately is really
just a coincidence, since neither the self energy nor the vertex function
is reproduced well at that value of the interaction strength (see Figures 
6--8).  There is
a very delicate balance between the self-energy terms (which tend to reduce
$T_c$ as the interaction increases) and the vertex-function terms (which
tend to increase $T_c$ as the interaction increases) that causes $T_c$ to
have a peak for the IPT approximation, but continue to increase for
the approximations in this contribution.

We end with a calculation of the phase diagrams off of half-filling, which
show the transition from the
checkerboard CDW insulator to a SC.  These results are plotted
in Figure~10.  The horizontal axis is the electron filling and the vertical axis
is the transition temperature to either a CDW or SC phase.  The CDW-SC
phase boundary lies at the points where the phase diagrams have a slope
discontinuity.  Three values of interaction strength are included, and both
the approximate solutions and the QMC results are presented (the lines through
the QMC data are just a guide to the eye).  The agreement for
$T_c$ is quite good for the weak-coupling value $g=0.4$, with all approximations
going through the QMC data for the SC phase, and overestimating $T_c$ 
only slightly for the CDW phase.  Here, the Hartree-Fock expansion is 
actually worse than the Hartree expansion for the CDW phase.  The 
calculation also seems to underestimate the location of the CDW-SC phase 
boundary.  The case with $g=0.5$ begins to show how the solutions develop 
qualitatively wrong behavior near half filling.  All but the  Hartree-Fock
expansion show a suppression of $T_c$ as the filling approaches $1$.  Such
a suppression was never seen in the QMC simulations, and is likely an artifact 
of the non-conserving nature of the approximations\cite{freericks_ipt}.  
It continues to be
clear that the SC solutions are approximated better than the CDW solutions,
and that the approximations behave worst near half-filling.  In particular,
it is important to note that the simple n-consistent scheme, which worked
so well for the Hubbard model, does not appear to work as well for the
electron-phonon problem examined
here, because the phase diagrams (H and RH) continue to have
qualitatively incorrect behavior.  Finally we examine the strongly coupled case 
$g=0.625$.  Here the renormalized expansions do a better job at
repairing the qualitatively wrong behavior of a $T_c$ suppression near
half-filling, but they do so by greatly enhancing the $T_c$'s at and
near half-filling from the truncated expansions.  It is clear, that 
at this large a value of the coupling strength, none of these weak-coupling
approaches is doing a good job at predicting the phase diagrams.
It is interesting to note that the close agreement of the Hartree
and renormalized Hartree calculations shows that the majority of the shift of
the chemical potential comes from the Fock dressing of the Hartree diagram.
Surprisingly, the inclusion of this diagram has a large effect on the
phase diagram when $|U|>1$.

\section{Conclusions}
What are the conclusions that can be drawn from this work? Even at a weak
value of the bare interaction strength ($g=0.4$) both the self energy and the
irreducible vertex functions are not approximated accurately by the
perturbation theory.  This implies that third-order (and higher-order) diagrams
are important and cannot be neglected.  For example, it is possible that
a scheme that uses an expansion with dressed phonon propagators would be
more accurate, which is work in progress.  Nevertheless,
the perturbation theory does appear to produce a more accurate approximation
to the phase diagrams than the individual self-energies.  
On the other hand, these results show that there is no simple scheme that
will allow one to approximate the electron-phonon problem accurately for
all values of the phonon frequency.  So one should not necessarily rely on 
approximations that employ purely electronic models (such as the Hubbard
model) to carry over to the electron-phonon problem and work well when
the phonon frequency is small to moderate.  Instead, it is more fruitful to
work on generalizations of the Migdal-Eliashberg theory that work with dressed
phonons, but include higher-order nonadiabatic effects such as vertex
corrections.  This should save a lot of time in formulating accurate
approximation methods for real materials that have large enough phonon
energy scales that the effects of nonconstant density of states and
of vertex corrections cannot be neglected.
Working with dressed phonons actually makes a lot of sense for the 
electron-phonon problem, because experiments can directly measure the 
dressed phonon spectral function, so it is readily available for use
in real materials calculations. Furthermore, the dressed phonon propagator
can be extracted from the QMC simulation and used in the perturbation theory 
for the electrons, just as is done from experiments on real materials.
Work along these lines is currently in progress.

\acknowledgments

We would like to thank Paul Miller and Joe Serene for useful discussions.
J.K.F. and W.C. acknowledge support for the initial stages of this work from
the Office of Naval Research Young Investigator Program N000149610828 and
J.K.F. and V.Z. acknowledge support for the final stages of this work from
the National Science Foundation INT-9722782 and DMR-9627778.  M.J. acknowledges
support from the National Science Foundation DMR-9704021 and DMR-9357199,
and from the Ohio Supercomputer Center.

\begin{figure}[t]
\caption{Feynman diagrams included in the Hartree expansion.  The top
figure shows the Hartree diagram, which is incorporated into $G_{MF}$, and
the bottom shows the Yosida-Yamada functional about the Hartree mean-field 
solution expanded through second order.
The wiggly lines are bare phonon propagators, and the straight lines are
Hartree (mean-field) propagators.
The fifth diagram, that involves the Fock dressing of the Hartree diagram
was not included in the previous work.}
\end{figure}

\begin{figure}[t]
\caption{Feynman diagrams included in the Hartree-Fock expansion.  The top
figure shows the Hartree and Fock diagrams, which are incorporated into 
$G_{MF}$, and
the bottom shows the Yosida-Yamada functional about the Hartree-Fock
mean-field solution expanded through second order.
The wiggly lines are bare phonon propagators and the straight lines are 
Hartree-Fock (mean-field) propagators.}
\end{figure}

\begin{figure}[t]
\caption{Modified Yosida-Yamada functional for the renormalized expansions.
(a) The renormalized-Hartree expansion (also known as the n-consistent
approximation), in which all dressings of the Hartree diagram are removed from
the Yosida-Yamada functional. (b) The renormalized Hartree-Fock expansion,
in which all dressings of the Hartree and the Fock diagrams are removed
from the Yosida-Yamada functional. The wiggly lines are the phonon propagators
and the straight lines are the corresponding mean-field Green's functions
(Hartree or Hartree-Fock).}
\end{figure}

\begin{figure}[t]
\caption{Irreducible vertex functions for (a) charge-density-wave and
(b) superconducting order. The wiggly lines are the bare phonon propagators
and the straight lines are the corresponding mean-field propagators
(Hartree or Hartree-Fock).}
\end{figure}

\begin{figure}[t]
\caption{Iterative algorithm for solving the self-consistent perturbation
theory.  As described in the text, one starts from an initial self-energy,
then constructs the local Green's function, the effective-medium Green's 
function, and then the mean-field Green's function.  The self-energy is
then determined from the Yosida-Yamada expansion, and if the calculation
is not converged, then the chemical potential is updated to adjust the
electron-filling and the process is repeated.  If the calculation has converged,
then the irreducible vertex functions are determined from Figure 4, and
the largest eigenvalue of the scattering matrix is then calculated.}
\end{figure}

\begin{figure}[t]
\caption{Self energy and irreducible vertex functions for the Holstein model
at half filling ($n=1.0$) and weak-coupling $g=0.4$.  The temperature is
just above $T_c$ at $T=1/16$.  Figure 6(a) shows the quasiparticle 
renormalization factor minus 1 as a function of frequency.  The four different
expansion schemes (Hartree, Renormalized Hartree, Hartree-Fock, and
Renormalized Hartree-Fock) are plotted with lines, and the quantum
Monte Carlo is plotted with solid dots.  We believe the combined errors
on the simulation to be on the order of a few percent. Figure 6(b) is the CDW
vertex and Figure 6(c) is the SC vertex.}
\end{figure}

\begin{figure}[t]
\caption{Self energy and irreducible vertex functions for the Holstein model
at half filling ($n=1.0$) and moderate-coupling $g=0.5$.   The temperature is
just above $T_c$ at $T=1/9$.  Figures 7(a), (b), and (c) are the quasiparticle
renormalization factor minus 1, the CDW vertex, and the SC vertex 
respectively.  Note
how the agreement with the perturbation theory worsens as the coupling
strength is increased.}
\end{figure}

\begin{figure}[t]
\caption{Self energy and irreducible vertex functions for the Holstein model
at quarter filling ($n=0.5$) and weak-coupling $g=0.4$.  The temperature is
the same as in Figure 6, $T=1/16$. Figure 8(a) is the quasiparticle
renormalization factor minus 1 and Figure 8(b) is the SC vertex.  Note how
the perturbation theory has improved, but less than what would have been
expected naively.}
\end{figure}

\begin{figure}
\caption{Phase diagram of the Holstein model at half filling.  The four
different expansion schemes (lines) are compared to the quantum Monte
Carlo (dots).  Note how all perturbative approximations do not show a peak
in the CDW transition temperature at half filling, but rather continue to
increase.  The Hartree and Renormalized Hartree expansions are identical
until the solution at half filling (with chemical potential 
$\mu=U)$ becomes unstable for the RH expansion, and the 
Green's functions acquire real parts.}
\end{figure}

\begin{figure}
\caption{Phase diagram of the Holstein model away from half filling
for three different coupling strengths $g=0.4$, $g=0.5$, and $g=0.625$.
The lines through the QMC data points are a guide to the eye.
Note that the SC transition temperatures are approximated better than the CDW
transition temperatures, and that the phase diagrams are predicted more
accurately than would be expected from the individual self-energies or
vertex functions.  As the coupling strength is increased the $T_c's$ are
all enhanced significantly, and the location of the CDW-SC phase boundary
is predicted less accurately.  The renormalized expansions do sometimes
improve the qualitative shape of the phase boundaries, to show a maximum
$T_c$ at half filling, but the perturbative approach is definitely breaking
down as the coupling strength becomes larger than $0.5$.}
\end{figure}

\begin{figure}[tbp]
\epsfxsize=4.8in 
\epsffile{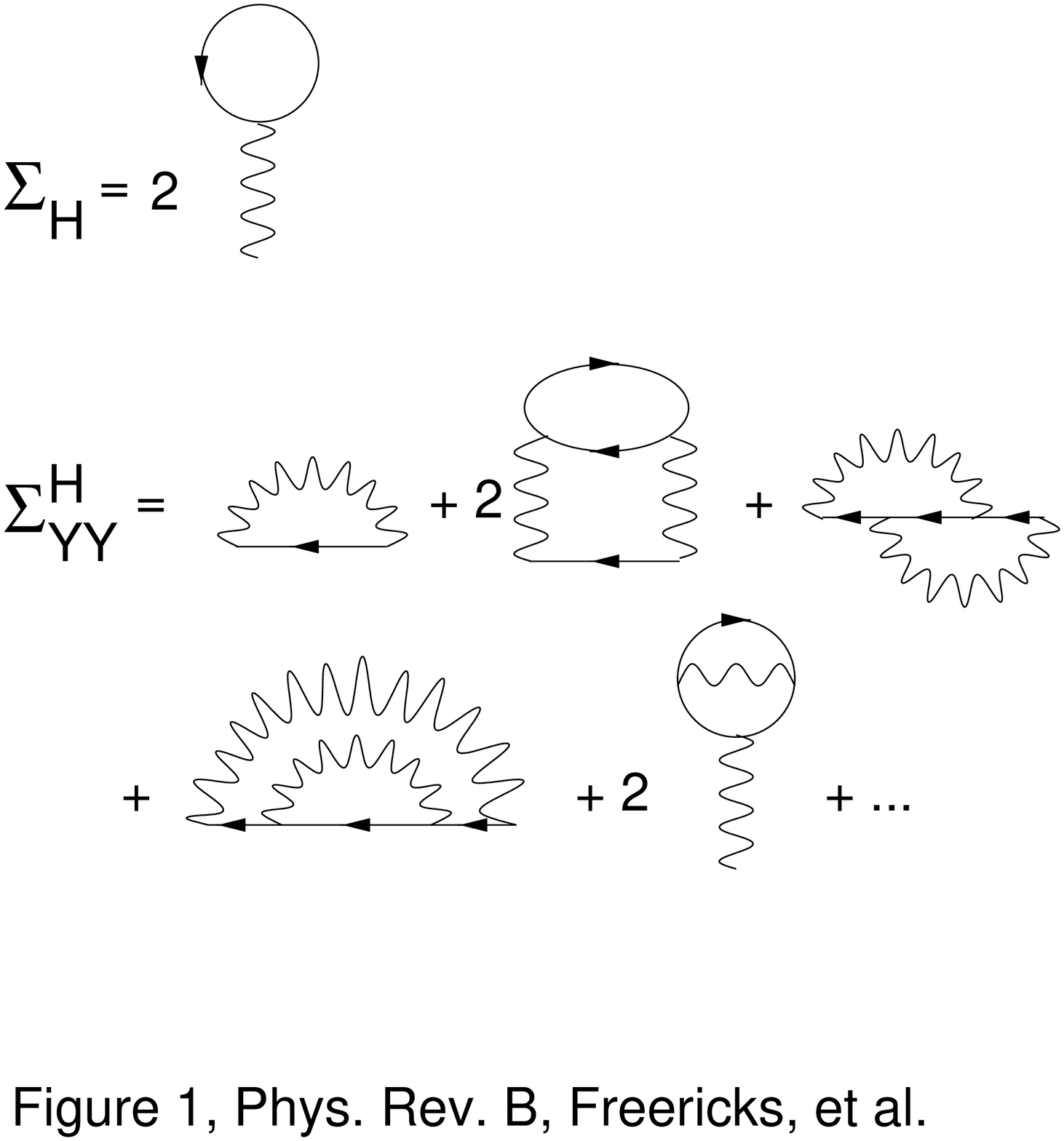}
\end{figure}

\begin{figure}[tbp]
\epsfxsize=3.9in 
\epsffile{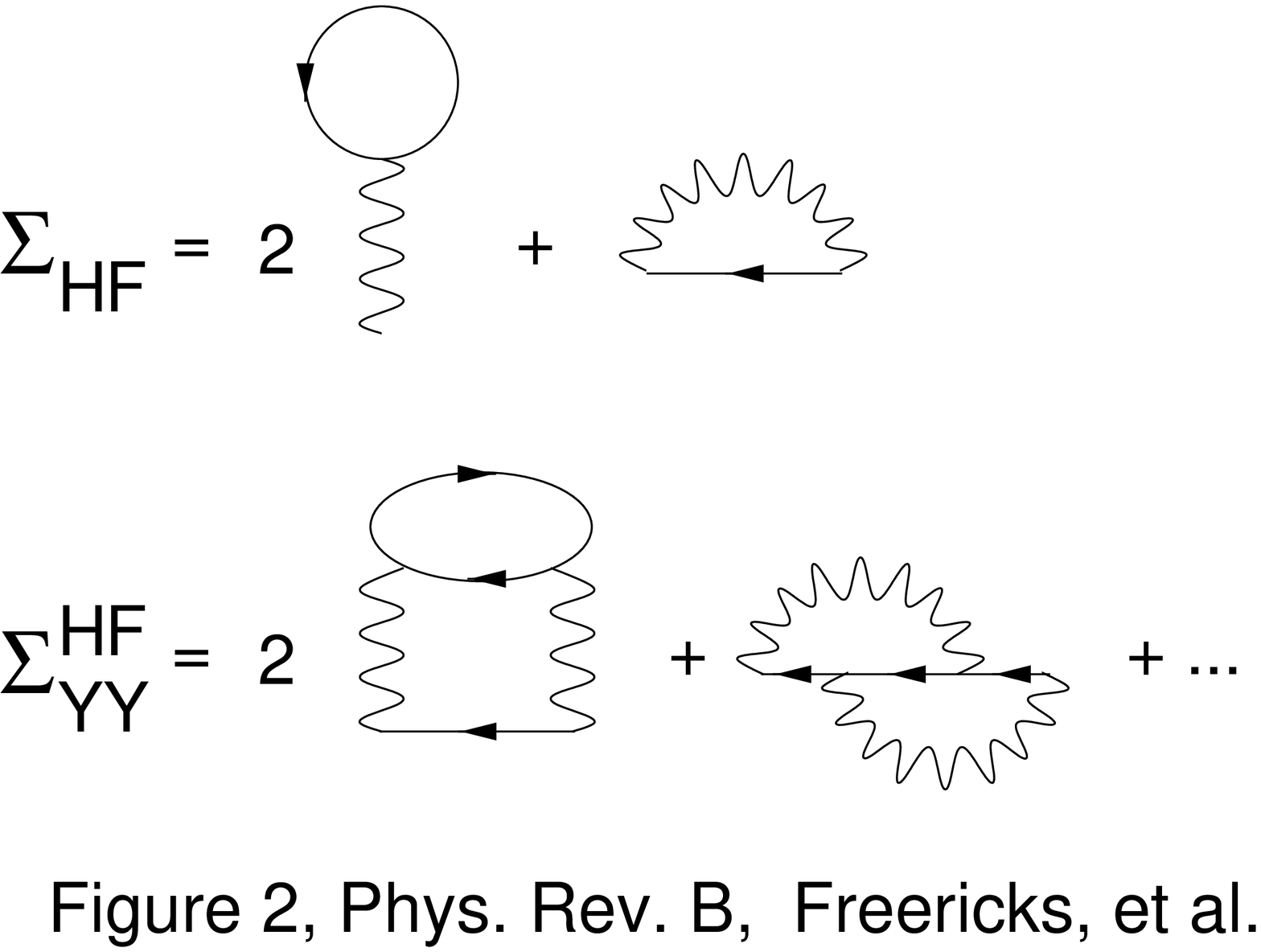}
\end{figure}

\begin{figure}[tbp]
\epsfxsize=7.0in 
\epsffile{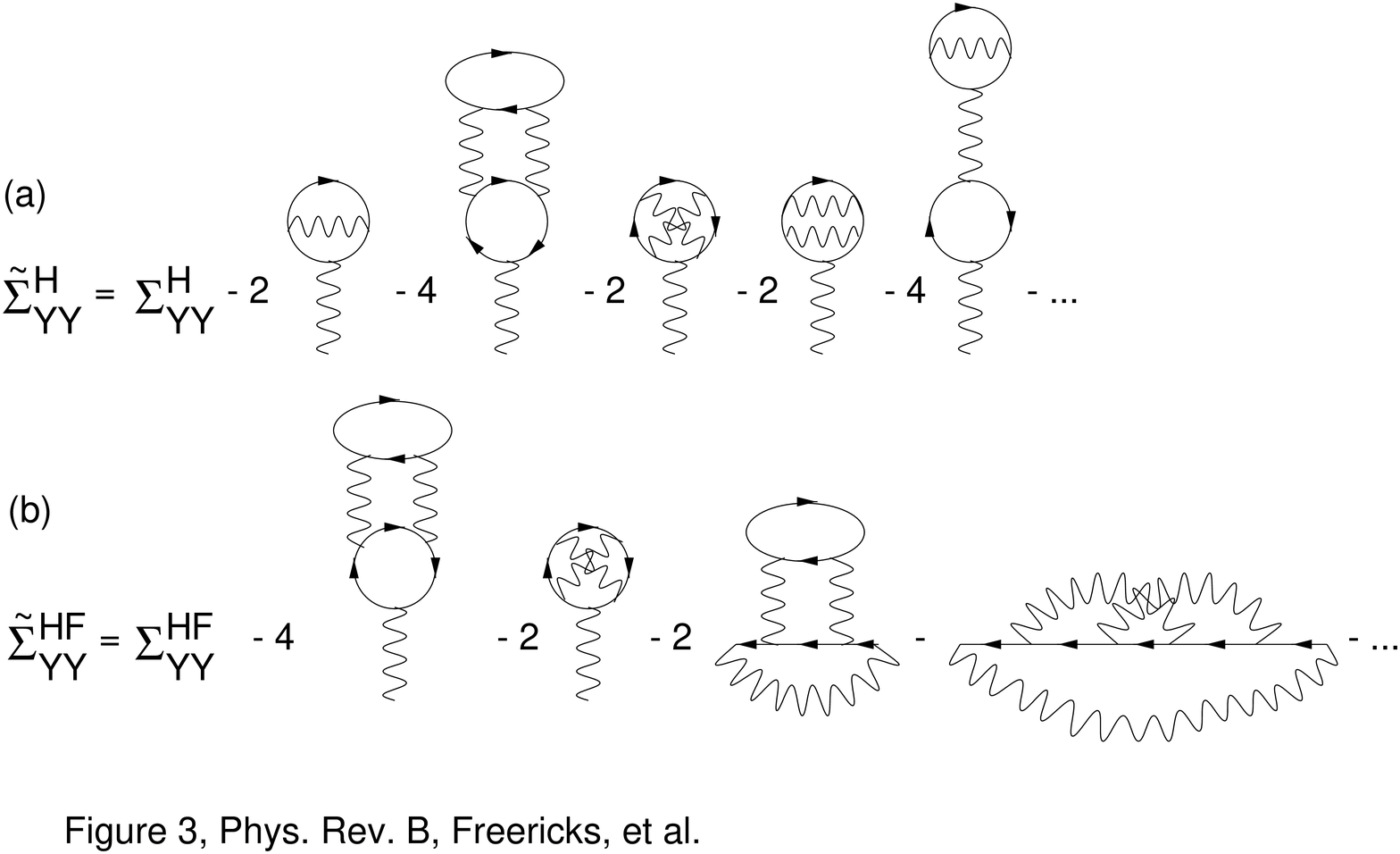}
\end{figure}

\begin{figure}[tbp]
\epsfxsize=6.0in 
\epsffile{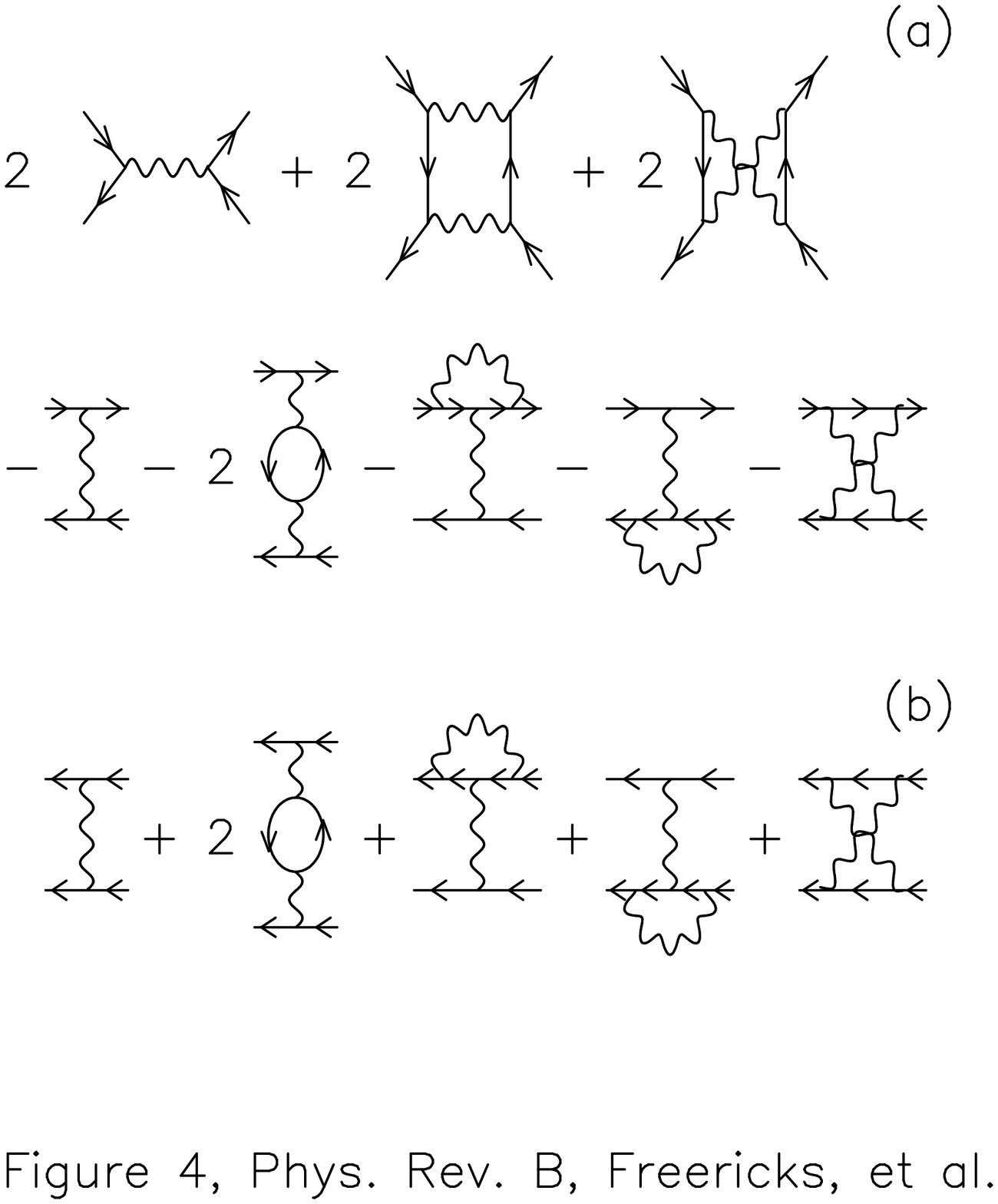}
\end{figure}

\begin{figure}[tbp]
\epsfxsize=6.5in 
\epsffile{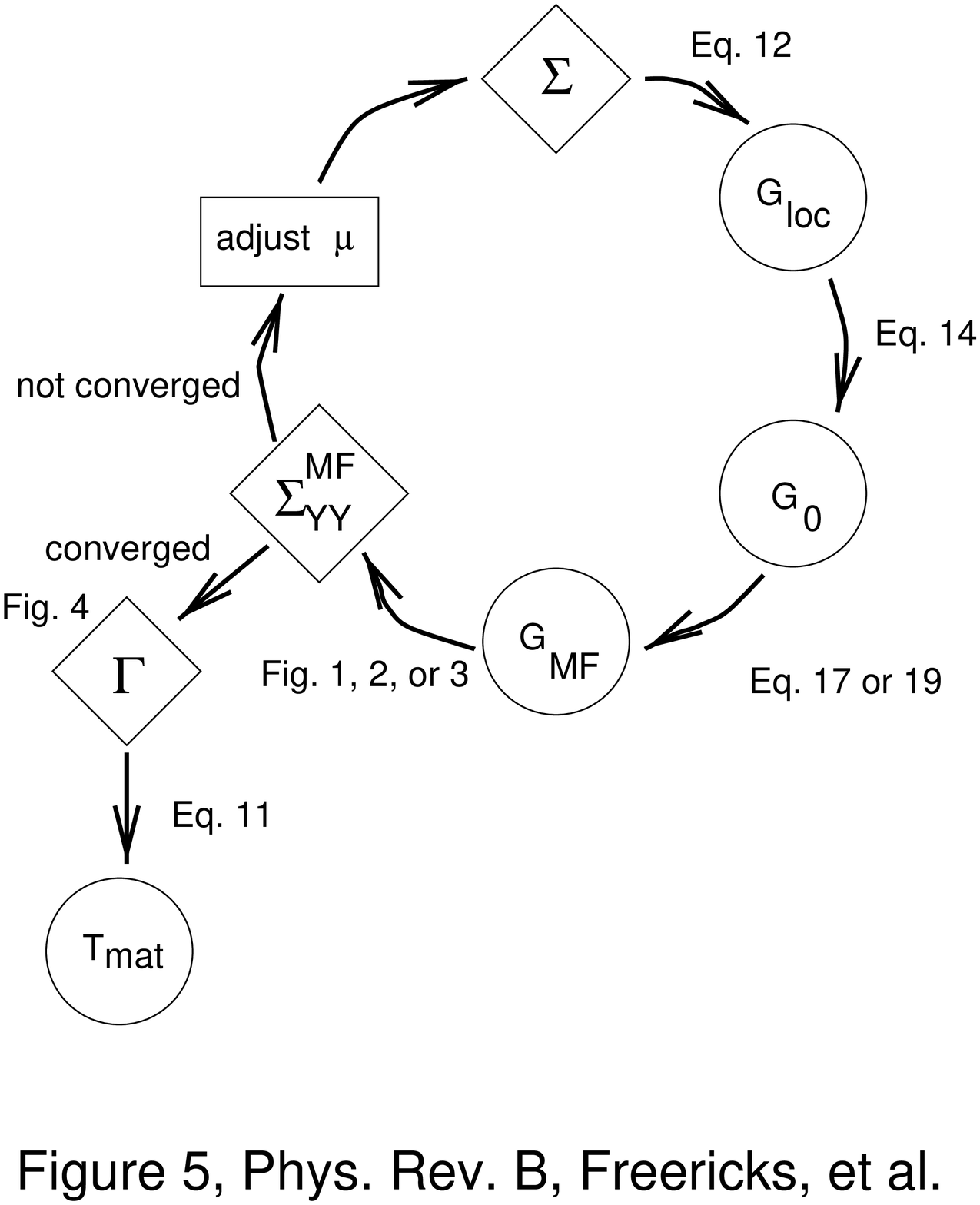}
\end{figure}

\begin{figure}[tbp]
\epsffile{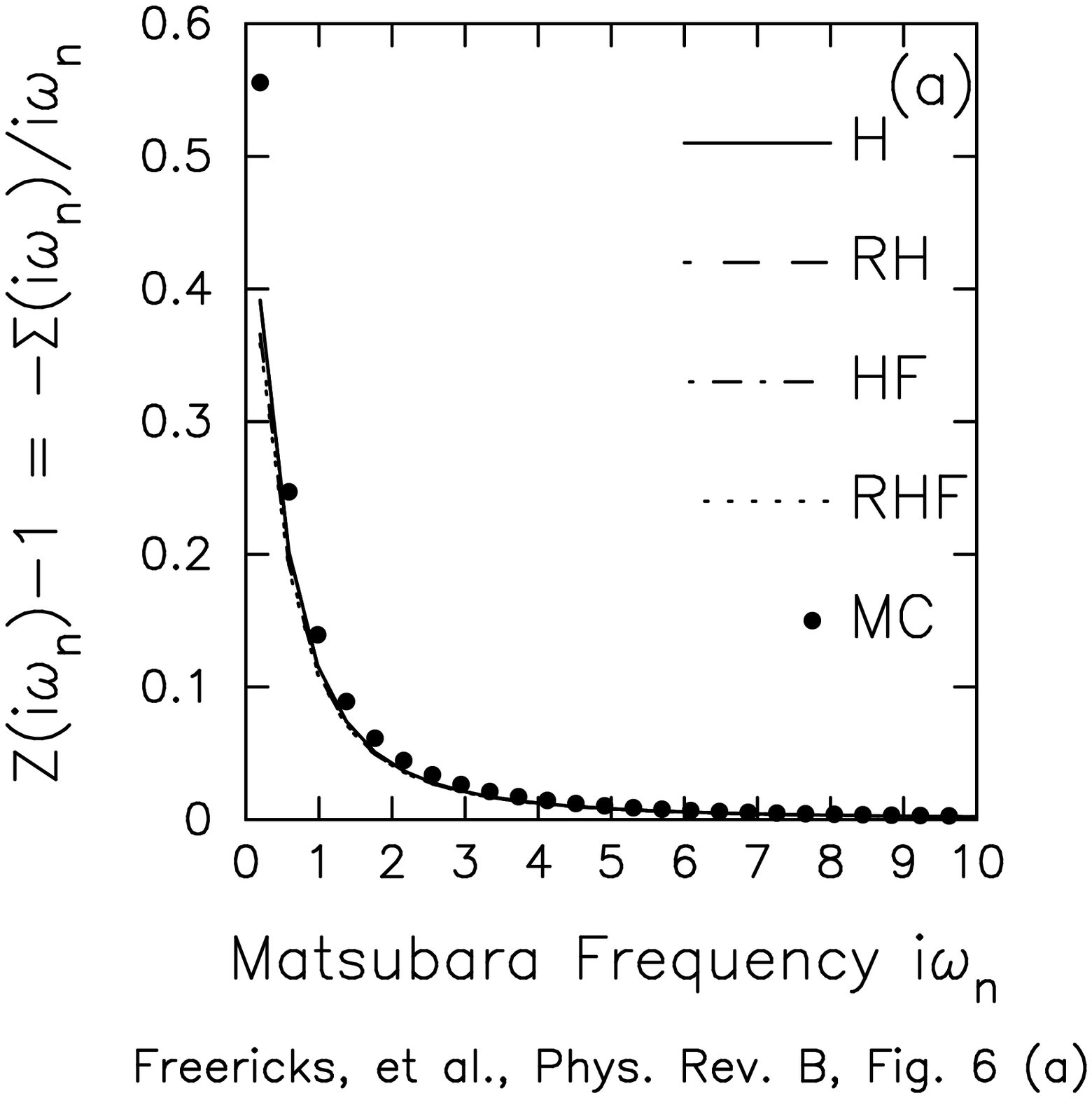}
\epsffile{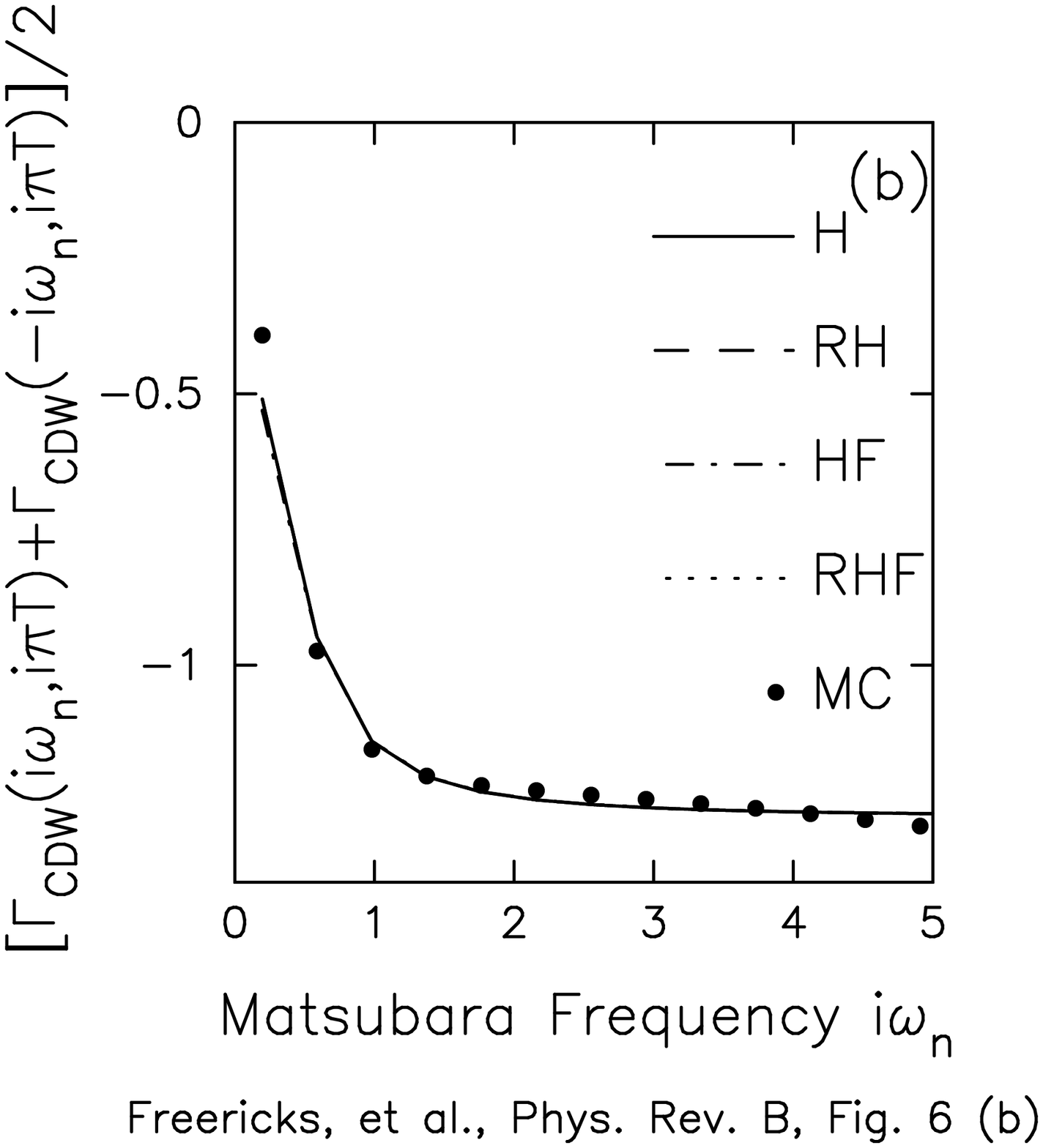}
\epsffile{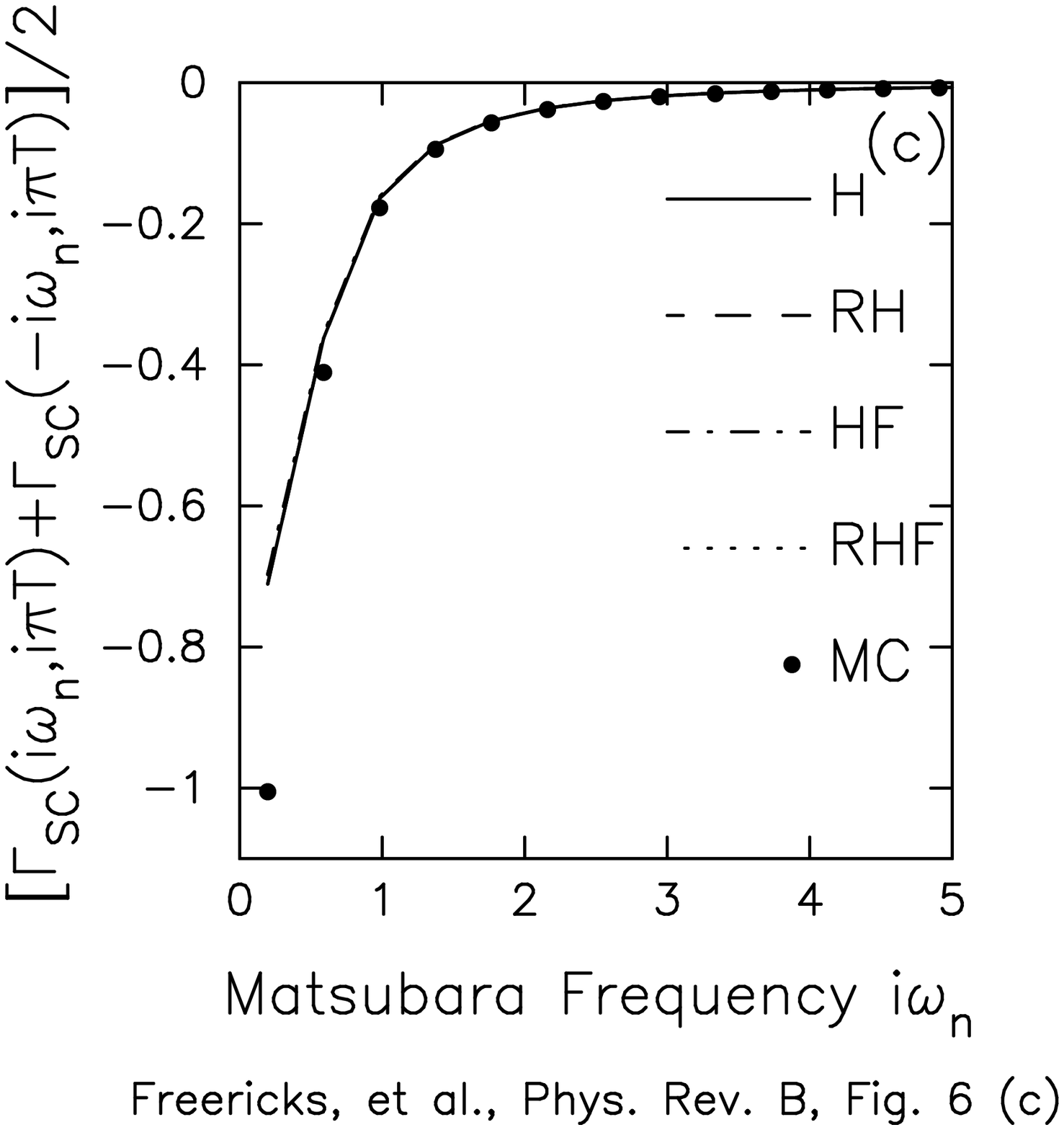}
\end{figure}

\begin{figure}[tbp]
\epsffile{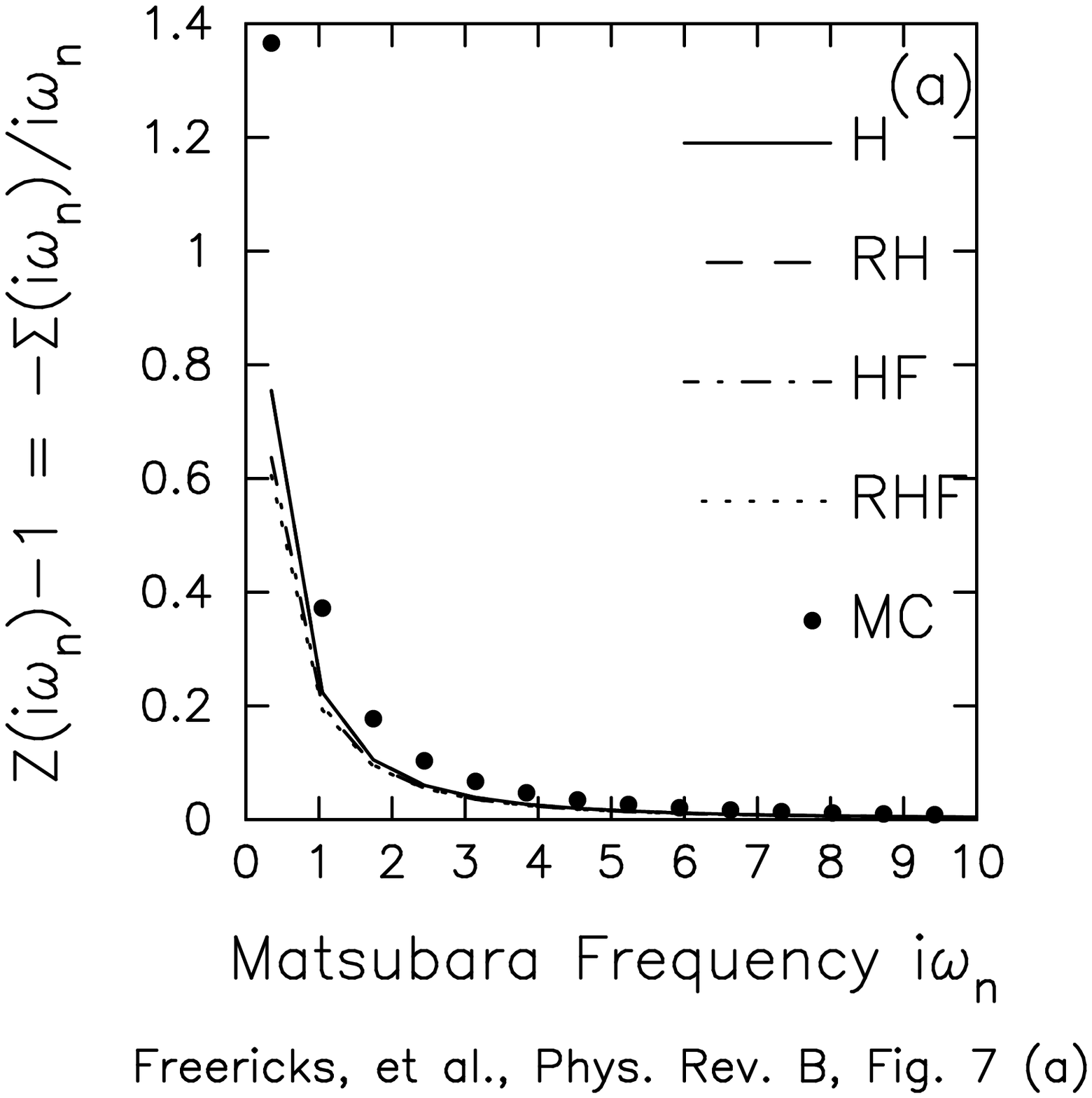}
\epsffile{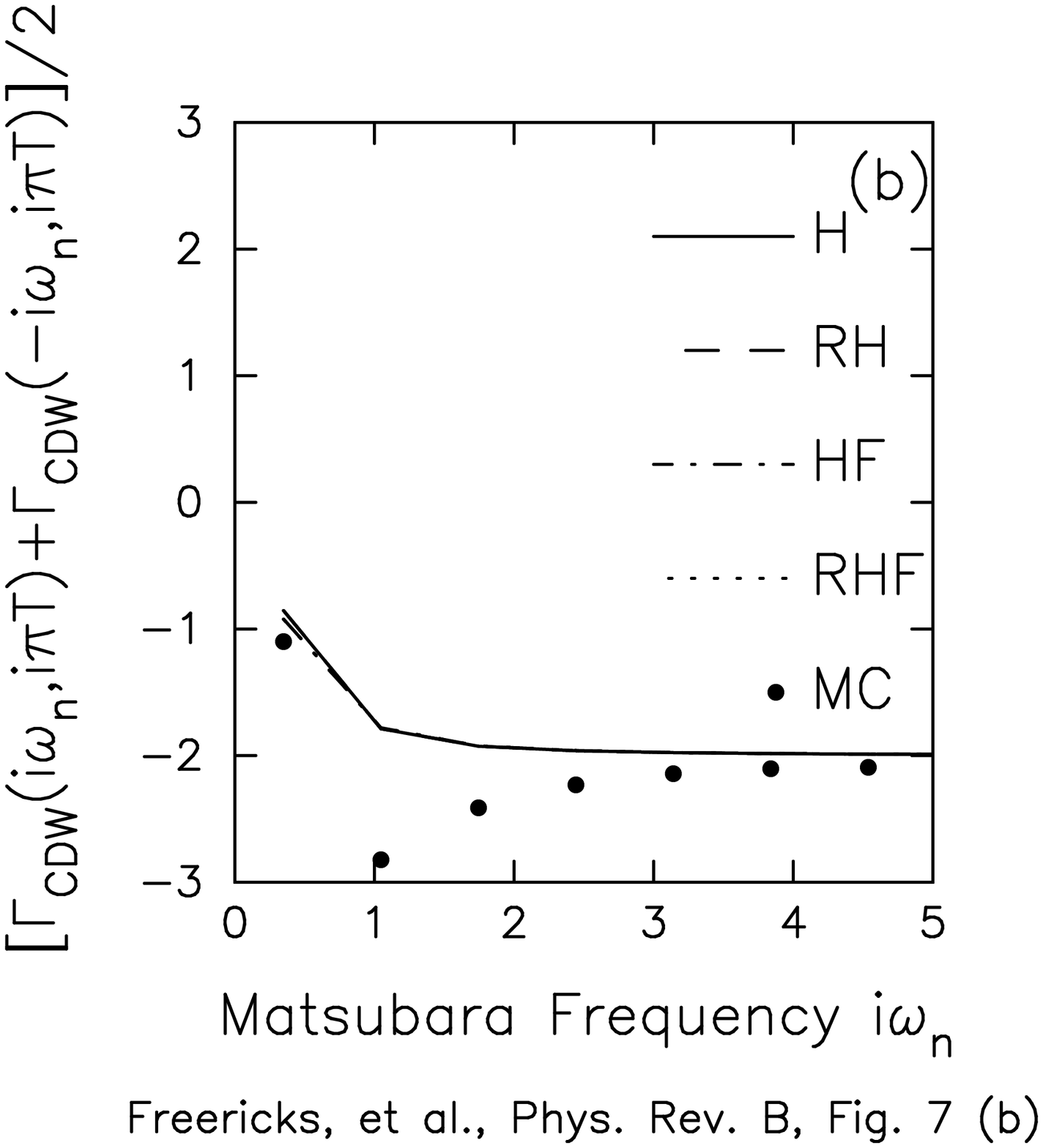}
\epsffile{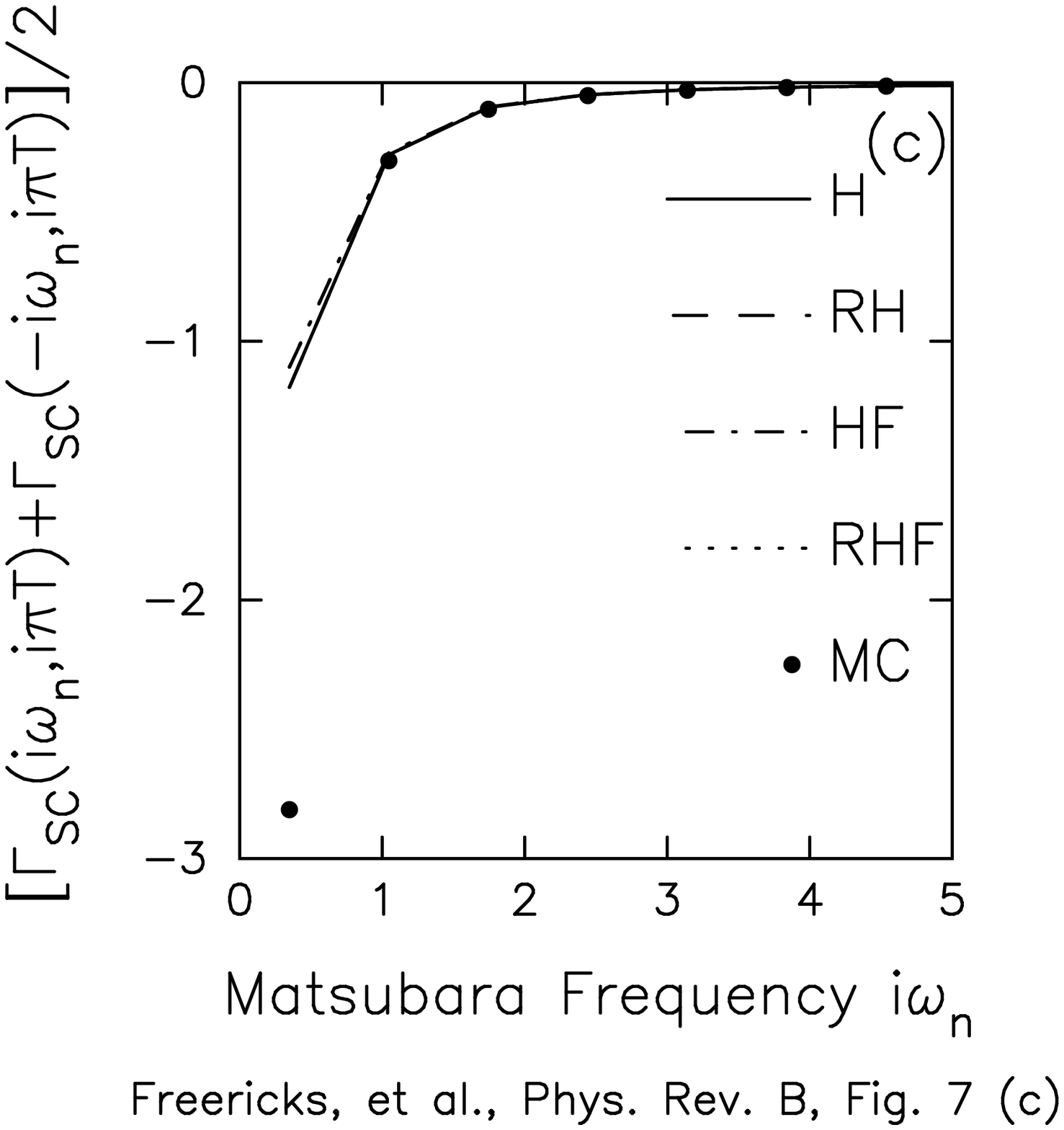}
\end{figure}

\begin{figure}[tbp]
\epsffile{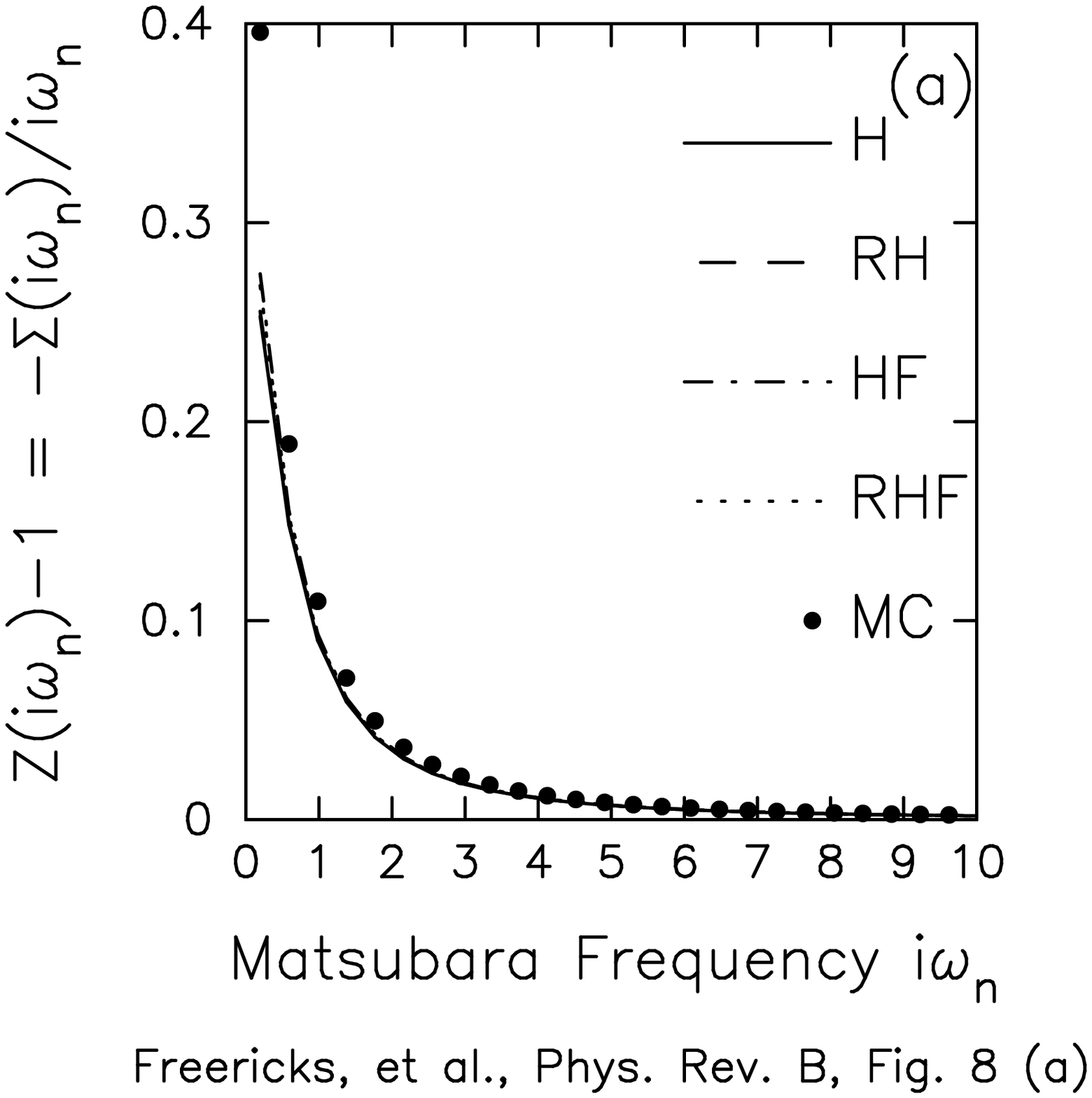}
\epsffile{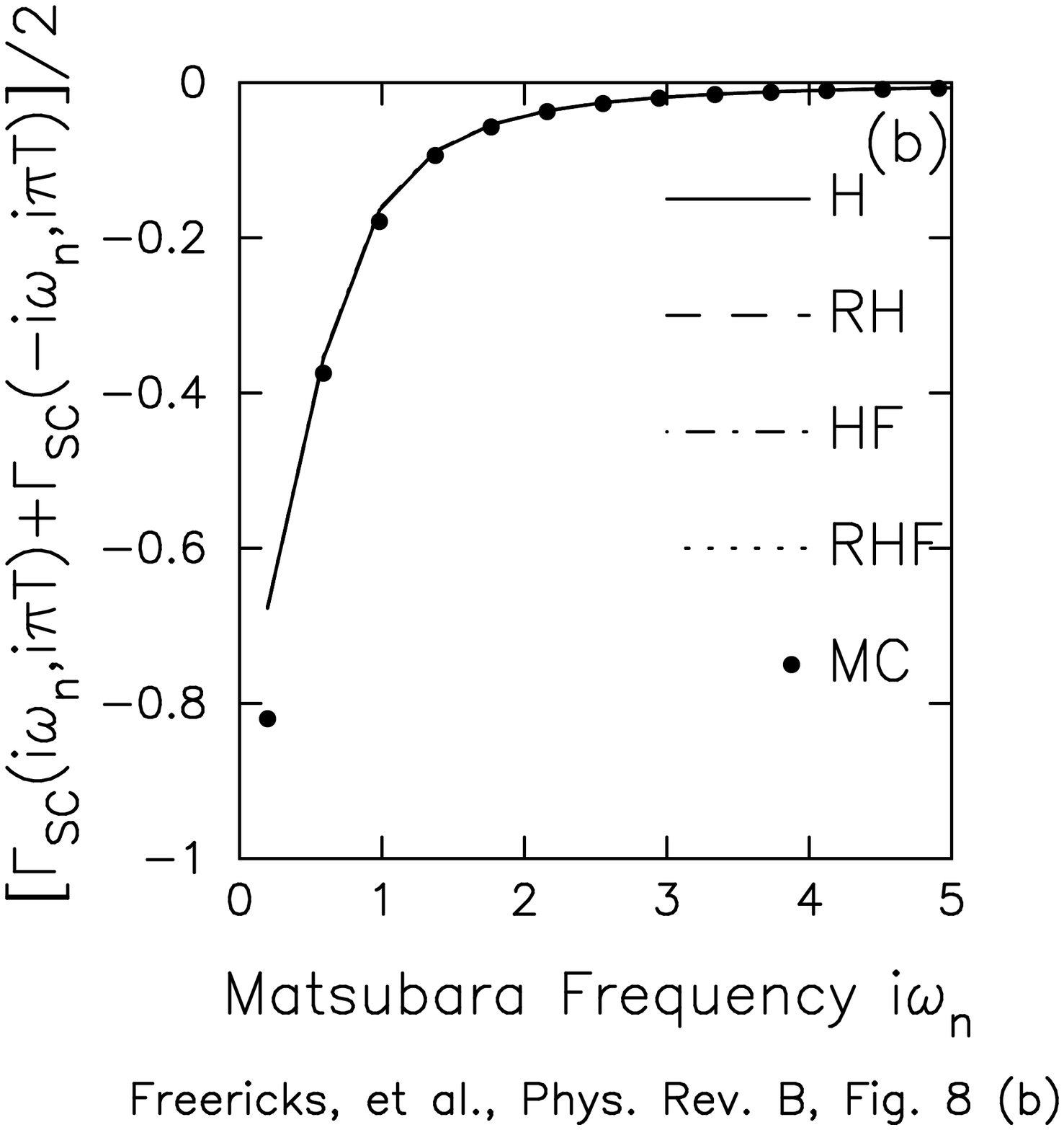}
\end{figure}

\begin{figure}[tbp]
\epsfxsize=5.5in
\epsffile{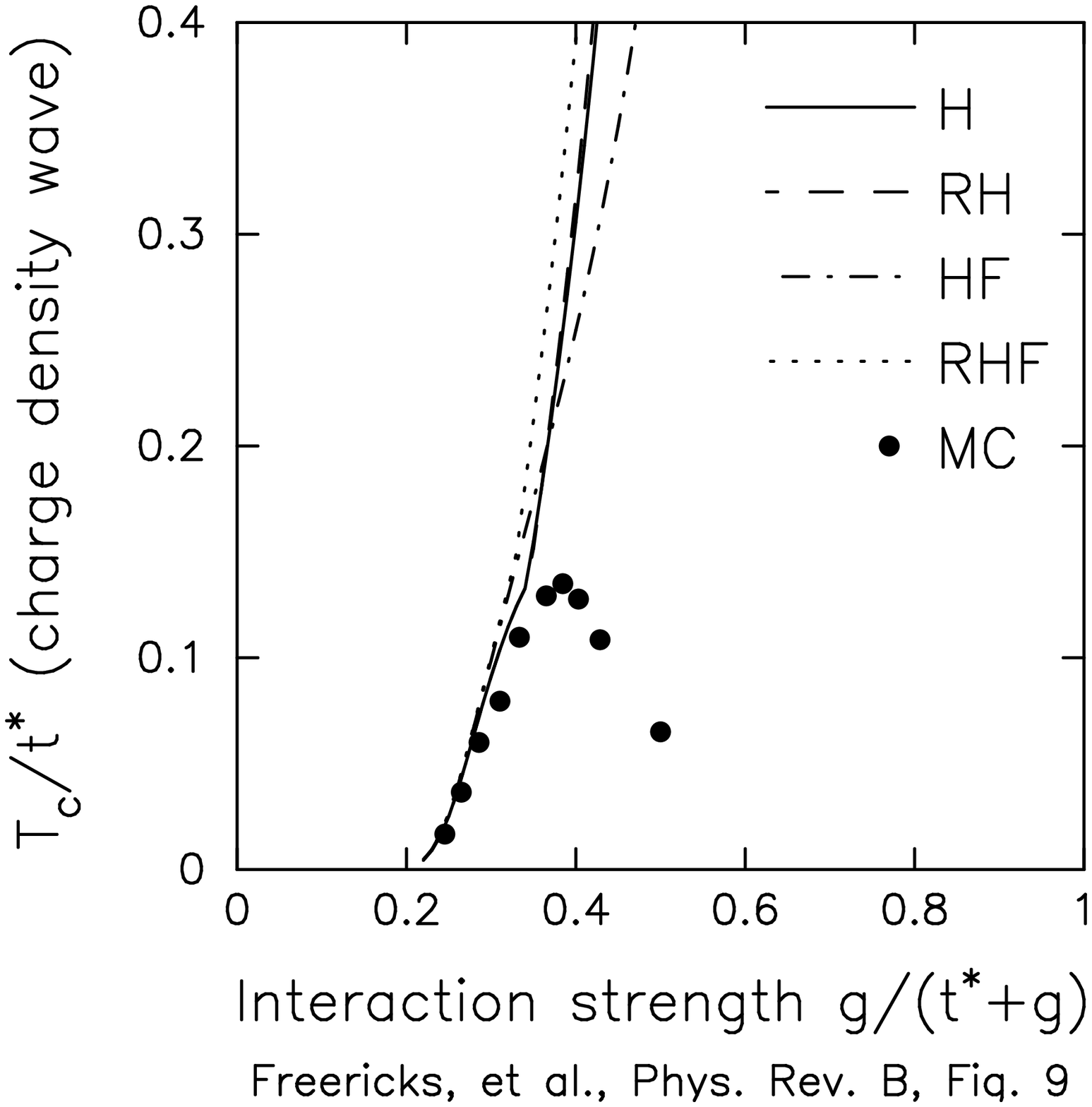}
\end{figure}

\begin{figure}[tbp]
\epsfxsize=5.5in
\epsffile{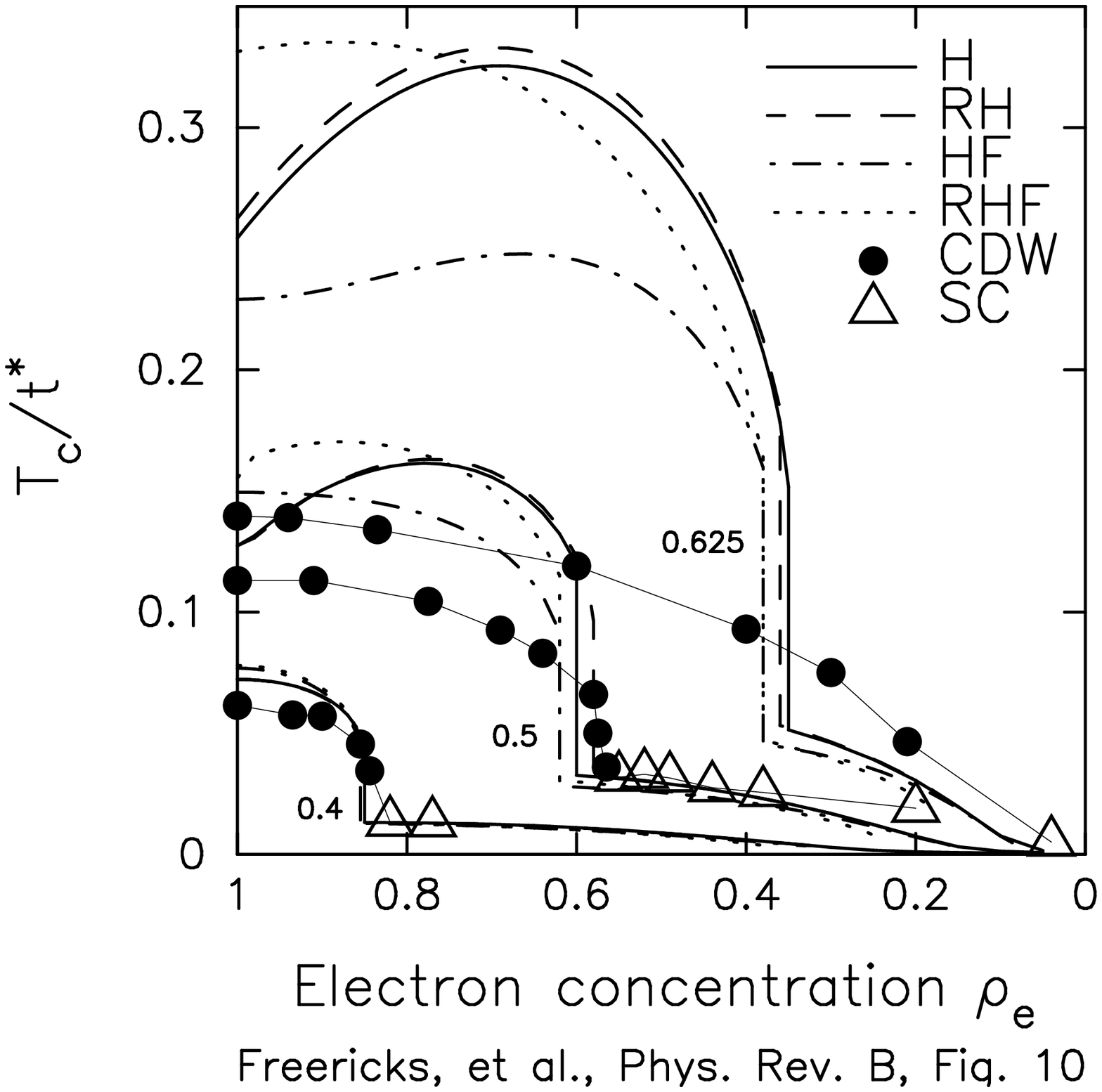}
\end{figure}


\begin{thebibliography}{10}

\bibitem{bcs} J. Bardeen, L. Cooper, and J. Schrieffer, Phys. Rev. {\bf 108},
1175 (1957).

\bibitem{migdal} A. B. Migdal, 
{Zh. Eksp. Teor. Fiz. {\bf 34}, 1438 (1958)}
{[Sov. Phys.-JETP {\bf 7}, 999 (1958)]}.

\bibitem{eliashberg} G. M. Eliashberg, 
{Zh. Eksp. Teor. Fiz. {\bf 38}, 966 (1960)}
{[Sov. Phys.-JETP {\bf 11}, 696 (1960)]}.

\bibitem{parks} {{\it Superconductivity}, edited by R. Parks 
(Marcel Dekker, Inc., New York, 1969)}.

\bibitem{reviews} P. B. Allen and R. C. Dynes, 
{Phys. Rev. B {\bf 12} 905 (1975)};
P. B. Allen and B. Mitrovi\' c, {Solid State Phys. {\bf 37}, 1 (1982)};
J. P. Carbotte, {Rev. Mod. Phys. {\bf 62}, 1027 (1990)};
H.-S. Wu, Z.-Y Weng, G. Ji, and Z.-F. Zhou,
{J. Phys. Chem. Solids {\bf 48}, 395 (1987)}.

\bibitem{holstein} T. Holstein, {Ann. Phys. {\bf 8}, 325 (1959)}.

\bibitem{hubbard} J. Hubbard, 
{Proc. Royal Soc. London (Ser. A) {\bf 276}, 238 (1963)}.

\bibitem{georges_kotliar} A. Georges and G. Kotliar, 
{ Phys. Rev. B {\bf 45}, 6479 (1992)}.

\bibitem{freericks_ipt} J. K. Freericks and M. Jarrell, Phys. Rev. B
{\bf 50}, 6939 (1994).

\bibitem{metzner_vollhardt} W. Metzner and D. Vollhardt, 
{Phys. Rev. Lett. {\bf 62}, 324 (1989)}.

\bibitem{schweitzer_czycholl} H. Schweitzer and G. Czycholl, 
{ Z. Phys. B {\bf 77}, 327 (1990)}.

\bibitem{metzner} W. Metzner, 
{ Phys. Rev. B {\bf 43}, 8549  (1991)}.

\bibitem{qmc_holstein} J. E. Hirsch and R. M. Fye, 
{Phys. Rev. Lett. {\bf 56}, 2521 (1986)},M. Jarrell,
{Phys. Rev. Lett. {\bf 69}, 168 (1992)};
M. Jarrell and Th. Pruschke,
{Z. Phys. B {\bf 90}, 187 (1993)};
Th. Pruschke, D. L. Cox, and M. Jarrell,
{Phys. Rev. B {\bf 47}, 3553 (1993)};
J. K. Freericks, M. Jarrell, 
and D. J. Scalapino,
{Phys. Rev. B {\bf 48}, 6302 (1993)};
{ Europhys. Lett. {\bf 25}, 37 (1994)}; J. K. Freericks and M. 
Jarrell,
in {\it Computer Simulations in Condensed Matter Physics}, Vol. VII, edited
by D. P. Landau, K. K. Mon, and H.-B. Sch\"uttler (Springer-Verlag, Heidelberg,
Berlin, 1994).

\bibitem{zlatic_horvatic_infd}  V. Zlati\'c and B. Horvati\'c, 
{Solid State Commun. {\bf 75}, 263 (1990)}.

\bibitem{brandt_mielsch} U. Brandt and C. Mielsch, 
{Z. Phys. B {\bf 75}, 365 (1989)}

\bibitem{muellerhartmann} E. M\"uller-Hartmann,
{Z. Phys. B {\bf 74}, 507 (1989)};
{Z. Phys. B {\bf 76}, 211 (1989)}; B. Menge and E. M\"uller-Hartmann,
{Z. Phys. B {\bf 82},237 (1991)}.

\bibitem{owen_scalapino} C. S. Owen and D. J. Scalapino,
{Physica {\bf 55}, 691 (1971)}.

\bibitem{wolff} P. W. Wolff, Phys.Rev. {\bf 124}, 1030 (1961);
H. R. Krishnamurty, J. W. Wilkins, and K. G. Wilson, Phys. Rev. B{\bf 21}, 
1003 (1980).

\bibitem{baym_kadanoff} G. Baym and L. P. Kadanoff, 
{Phys. Rev. {\bf 124}, 287 (1961)}; G. Baym, 
{Phys. Rev. {\bf 127}, 1391 (1962)}; N. E. Bickers and D. J. Scalapino,
{Ann. Phys. {\bf 193}, 206 (1989)}.

\bibitem{yosida_yamada} K. Yosida and K. Yamada,
{Prog. Theor. Phys. {\bf 46}, 244 (1970)}; K. Yamada,
{Prog. Theor. Phys. {\bf 53}, 970 (1975)}; K. Yosida and K. Yamada,
{Prog. Theor. Phys. {\bf 53}, 1286 (1975)}; K. Yamada,
{Prog. Theor. Phys. {\bf 54}, 316 (1975)}; 
{Prog. Theor. Phys. {\bf 55}, 1345 (1976)};
{Prog. Theor. Phys. {\bf 62}, 354 (1979)}.

\bibitem{zlatic.89} V. Zlati\'c and B. Horvati\'c,
{Phys. Rev. B {\bf },  (1989)}.
 
\bibitem{horvatic.85} B. Horvati\'c and V. Zlati\'c,
{Sol. St. Commun. {\bf 54}, 957 (1985)}.

\bibitem{n_consistent} V. Zlati\'c (unpublished).

\bibitem{kajuter_kotliar} H. Kajueter and G. Kotliar, Phys. Rev. Lett. {\bf 77},
131 (1996).

\end{thebibliography}
\end{document}